\documentclass[a4paper,12pt]{article}

\usepackage{amsmath,amsthm,amsfonts,amssymb,epsfig}
\usepackage{slashed}
\numberwithin{equation}{section}
\usepackage{dsfont}
\usepackage{cancel}
\usepackage{cite}

\usepackage{color}

\def\nn{\nonumber}

\def\tr{\mathrm{Tr}}

\def\beq{\begin{equation}}
\def\eeq{\end{equation}}
\def\be{\begin{equation}}
\def\ee{\end{equation}}
\def\bea{\begin{eqnarray}}
\def\eea{\end{eqnarray}}
\def\bal{\begin{align}}
\def\eal{\end{align}}

\def\2b2[#1,#2][#3,#4]{\left( \begin{array}{cc} #1 & #2 \\ #3 & #4 \end{array}
\right)}
\def\3b3[#1,#2,#3][#4,#5,#6][#7,#8,#9]{\left( \begin{array}{ccc} #1 & #2 #3 \\
#4 & #5 & #6\\#7&#8&#9\end{array} \right)}

\def\Ca[#1,#2]{C_{2\,#1}^{\ \ #2}}
\def\Y[#1,#2]{Y_{2\,#1}^{\ \ #2}}

\newcommand{\C}[1]{\mathcal{#1}}

\def\ov{\overline}

\author{Mark~D.~Goodsell\footnote{mark.goodsell@cern.ch}}
\date{}
\title{\vspace{-3cm}\small{
\hfill{CERN-PH-TH/2012-179}}\\[2cm]
\huge{Two-loop RGEs with Dirac gaugino masses}
}
\begin{document}
\maketitle
\vspace{-1cm}
\begin{center}
\emph{CERN, Theory Division, CH-1211 Geneva 23, Switzerland}
\end{center}
\abstract{The set of renormalisation group equations to two-loop order for general supersymmetric theories broken by soft and supersoft operators is completed. As an example, the explicit expressions for the RGEs in a Dirac gaugino extension of the (N)MSSM are presented.}

\section{Introduction}

Models with Dirac gaugino masses are attractive for a number of reasons. From a top-down model-building perspective, this is because they preserve R-symmetry, and so allow for simple supersymmetry-breaking sectors. This has attracted much interest in the literature \cite{Fayet:1978qc,Polchinski:1982an,Hall:1990hq,Fox:2002bu,Nelson:2002ca,Antoniadis:2005em,Antoniadis:2006uj,Hsieh:2007wq,Amigo:2008rc,Benakli:2008pg,Belanger:2009wf,Benakli:2009mk,Chun:2009zx,Benakli:2010gi,Carpenter:2010as,Kribs:2010md,Chun:2010hz,Choi:2010an,Benakli:2011kz,Abel:2011dc,Kumar:2011np,Benakli:2011vb,Davies:2011mp,Davies:2011js,Heikinheimo:2011fk,Rehermann:2011ax,Kribs:2012gx,Davies:2012vu,Argurio:2012cd}. On the other hand, if gauginos are found at the LHC, it must be determined whether they are of  Majorana or Dirac type \cite{Choi:2009ue,Choi:2010gc,Kalinowski:2011zz,Heikinheimo:2011fk,GoncalvesNetto:2012nt}. Moreover, with the current results from the LHC, Dirac mass terms allow the preservation of naturalness by having lower bounds on the gluino and squark masses in the Majorana case (they are ``super-safe'' \cite{Kribs:2012gx}); by increasing the Higgs mass due to additional couplings \cite{Fox:2002bu,Nelson:2002ca,Antoniadis:2006uj,Benakli:2011kz}; and because they do not cause the Higgs masses to run as strongly \cite{Fox:2002bu,Benakli:2010gi,Papucci:2011wy}. 

Particularly out of a desire to study the naturalness and Higgs sectors of such theories, it is important to know the renormalisation group equations (RGEs) for them.  The purpose of this paper is to complete the set of RGEs to two-loop order.

The standard soft supersymmetry-breaking mass terms are well known:
\begin{align}
\C{L}_{\rm standard} =& - \frac{1}{2} (m^2)^i_j \phi_i \phi^j - \frac{1}{2} M \lambda_A \lambda_A - \frac{1}{2} B^{ij} \phi_i \phi_j - \frac{1}{6} A^{ijk} \phi_i \phi_j \phi_k + {\rm h.c.}
\label{EQ:STANDARD}\end{align}
where the $\phi_i$ are the scalars of chiral multiplets $\Phi_i = \phi_i + \sqrt{2} (\theta \psi_i) + ...$ and $\phi^i \equiv (\phi_{i})^*$; $\lambda_A$ are gauginos. The above includes  Majorana gaugino masses $M$. However, Dirac gaugino masses fall under the category of ``non-standard'' soft terms:
\begin{align}
\C{L}_{\rm non-standard} =& - t^i \phi_i - m_D^{iA} \psi_i \lambda_A - \frac{1}{2} r^{ij}_{\ \ k}  \phi_i \phi_j \phi^k + {\rm h.c.}
\label{EQ:NONSTANDARD}\end{align} 
General choices of these terms will lead to quadratic divergences in singlet tadpoles, and this has often led to the terms being neglected or only included in theories without singlets. However, to give a Dirac mass to the Bino, a singlet superfield must be included, so it is necessary to worry about this issue. On the other hand, when supersymmetry is spontaneously broken, the expectation is that no quadratic divergences should be generated, and indeed it is generically found that only \emph{supersoft} \cite{Fox:2002bu} operators are generated:
\begin{align}
\C{L}_{\rm supersoft} \supset& \int {\rm d}^2 \theta \sqrt{2} m_D^{iA} \theta^\alpha \Phi_i W_{A\alpha}   + {\rm h.c.} \nn\\
\supset& - m_D^{iA} \psi_i \lambda_A + \sqrt{2} m_D^{iA} \phi_i D_A + {\rm h.c.},
\end{align} 
where $D_A$ is the D-term of the gauge group to which the adjoint couples with adjoint index `$A$'. They lead to a particular structure of non-standard soft terms such that the quadratic divergences exactly cancel. These interestingly augment supersymmetric trilinear couplings; including a superpotential
\begin{align}
W =& L^i \Phi_i + \frac{1}{2} \mu^{ij} \Phi_i \Phi_j + \frac{1}{6} Y^{ijk} \Phi_i \Phi_j \Phi_k
\label{EQ:SUPERPOTENTIAL}\end{align}
the non-standard trilinear couplings are 
\begin{align}
r^{ij}_{\ \ k} =&Y^{ijm} \mu_{mk} + \sqrt{2} g ( m_D^{iA} (t^A)^j_k + m_D^{jA} (t^A)^i_k) \nn\\
r_{ij}^{\ \ k} =&Y_{ijm} \mu^{mk} + \sqrt{2} g ( m_{DiA} (t^A)^k_j +  m_{DjA} (t^A)^k_i)
\label{EQ:RTERMS}\end{align} 
where $(t^A)^i_j$ are the generators of the gauge group. Of course, the supersymmetric terms do not generate quadratic divergences -- these are cancelled by fermion loops -- and  only receive wavefunction renormalisation. The would-be quadratic divergences from the non-supersymmetric, supersoft, piece are not cancelled by fermion loops but instead vanish when they are all summed.  

If there is a Fayet--Iliopoulos (FI) term $\xi_Y$ (where $Y$ denotes a $U(1)$ index) then a contribution to the tadpole is generated of
\begin{align}
\Delta t^a = & \sqrt{2} m_D^{a Y} \xi_Y.
\label{EQ:FITADPOLE}\end{align}
Typically, however, any FI term generated can be absorbed into the soft masses; this equation shows that in the presence of Dirac gaugino mass terms it should also be absorbed into a shift of the tadpole. Interestingly, there is also a supersymmetric term that emulates a tadpole equal to $\C{L} \supset -\mu^{ij} L_j \phi_i + c.c.$; of course this only has wavefunction renormalisation, just as for the trilinear terms above.

An important point is that the supersoft operator also generates contributions to the standard soft breaking terms via the self-coupling: from integrating out the auxiliary $D$-field there are terms
\begin{align}
\C{L} \supset& - (m_D^{iA} \phi_i + m_{DiA} \phi^i) (m_D^{jA} \phi_j + m_{DjA} \phi^j)
\end{align}
and so 
\begin{align}
\Delta (m^2)^i_j =& 2 m_D^{iA}m_{DjA}  \equiv 2 (m_D^2)^i_j \nn\\
\Delta B^{ij} =& 2 m_D^{iA}m_D^{jA}  \equiv 2 (m_D^2)^{ij}.
\label{EQ:SHIFTS}\end{align}

The two-loop renormalisation group equations (RGEs) for standard SUSY-breaking terms were derived some time ago \cite{Jack:1994kd,Martin:1993zk,Jack:2001ew}, and later the generic RGEs for non-standard soft terms were calculated \emph{in the absence of singlets} \cite{Jack:1999ud,Jack:1999fa}. They also found that restricting the non-standard terms to only be generated from supersoft operators defined a renormalisation group invariant trajectory, explained by the holomorphic nature of the supersoft operator. This means that equations (\ref{EQ:RTERMS}) and (\ref{EQ:SHIFTS}) are true at any renormalisation scale. The supersoft operator only obtains wavefunction renormalisation, so its beta-function is
\begin{align}
\beta_{m_D^{iA}} =& \gamma^i_j m_D^{jA} + \frac{\beta_g}{g}  m_D^{iA}
\end{align}
where $\gamma^i_j$ is the anomalous dimension of the adjoint superfield, and $\beta_g$ is the beta-function for the gauge coupling. Thus in a theory with Dirac gaugino mass terms, the RGEs for the standard soft terms can be found and evolved \emph{ignoring the Dirac gaugino mass}, and then at the scale of interest the shifts (\ref{EQ:SHIFTS}) can be applied to find the physical masses. In a theory without gauge singlets, this is then enough to determine all of the RGEs in the theory. However, when there are singlets -- such as when there is a Dirac mass for the Bino --  the RGE for the tadpole is also required, which is a non-standard term so may depend upon the Dirac gaugino mass, and not just via the above shifts. Considering that the singlet superfield may couple to the Higgs via a term 
\begin{align}
W \supset \lambda_S \mathbf{SH_u\!\cdot\!H_d}
\end{align}  
it is clear that knowing the size of the singlet tapdole (indeed, ensuring that it is not too large, since it is not protected by R symmetry for example) is vital in order to investigate electroweak symmetry breaking and determine the Higgs mass. The main result of this paper is to determine these RGEs to two-loop order. 

In section \ref{SEC:TADPOLE} the result is presented, with an explanation of how the different terms arise. The method used is that of  Martin and Vaughn \cite{Martin:1993zk}, deriving the RGE for a tadpole in a general renormalisable theory from the expressions given in \cite{Machacek:1983tz,Machacek:1983fi,Machacek:1984zw,Luo:2002ti}, then specialising to the softly broken supersymmetric case, transforming from $\ov{\rm MS}$ to $\ov{\rm DR}'$. To do this the rules given in \cite{Martin:1993yx} will be used, and augmented with a new rule for Dirac gaugino mass terms. In section \ref{SEC:MODEL} and the appendix the RGEs are derived for a minimal Dirac gaugino extension of the supersymmetric standard model with rather general couplings.

\section{Tadpole RGEs}
\label{SEC:TADPOLE}

There are several ways to derive the RGEs for softly broken supersymmetric models: by diagrams in component fields; by supergraphs; by RG invariance of the effective potential \cite{Jack:1999ud}; or by translating the results from a general renormalisable theory into the broken supersymmetric case. This last approach is the one adopted here, although the one-loop result was checked via the effective potential method. 

\subsection{Tadpole in non-supersymmetric theories}
\label{SEC:NONSUSY}

The first step in calculating the tadpole is to write down the expression in general non-supersymmetric theories. This can be derived using spurions from the RGEs for the quartic coupling in a general renormalisable theory given in\cite{Machacek:1983tz,Machacek:1983fi,Machacek:1984zw,Luo:2002ti}. Such a theory with real scalars $\phi_a$ and complex fermions $\psi_i$ has couplings
\begin{align}
\C{L} \supset& - \frac{1}{24} \lambda_{abcd} \phi_a \phi_b \phi_c \phi_d - \frac{1}{6} h_{abc} \phi_a \phi_b \phi_c- \frac{1}{2} m^2_{ab} \phi_a \phi_b  - t_{a} \phi_a \nn\\
&- \frac{1}{2}\bigg[ (m_{f})_{ij} (\psi_i \psi_j) + (Y^a)_{ij} \phi_a (\psi_i \psi_j) + h.c.\bigg]
\end{align}
in addition to a gauge coupling $g$.

The one-loop tadpole RGE is found to be
\begin{align}
(4\pi)^2 \beta_{t_a}=& 2\kappa Y_2(S)^{ab}  t_b + h_{aef} m^2_{ef} - 4 \kappa\tr(Y_a m_f^\dagger m_f m_f^\dagger)- 4 \kappa  \tr(Y_a^\dagger m_f m_f^\dagger m_f)
\end{align}
where $\kappa = 1/2$ for Weyl fermions (or $1$ for Dirac fermions) and
\begin{align}
Y_2(S)^{ab} \equiv&\frac{1}{2} \tr( Y^{\dagger a} Y^b + Y^{\dagger b} Y^a).
\end{align}

The two-loop tadpole RGE is 
\begin{align}
\beta_{t_a} =& (\gamma_S^{(2)})^b_a t_b \nn\\
&+ h_{aef} m^2_{eg} ( 8 g^2  C_2^{\ fg} -  4\kappa Y_2(S)^{fg}) - \frac{g^2}{2} h_{aef} h_{egh} h_{fgh} - g^2 m^2_{ef} \lambda_{aegh} h_{fgh} \nn\\
&+4\kappa \bigg(2 \ov{H}^\lambda_{a} +  H^Y_{a} + 2 \ov{H}^Y_{a} + 2 H^3_{a} -  g^2 H^F_{a} \bigg)
\end{align}
where now $ C_2^{\ fg}$ is the quadratic casimir of the gauge group for the representation carried by fields $f$ and $g$, and 
\begin{align}
(\gamma_S^{(2)})^b_a \equiv&  \frac{1}{24} \lambda_{acde} \lambda_{bcde} - \frac{3}{2}\kappa \tr(Y^a Y^{\dagger b}Y^c Y^{\dagger c}  )-\kappa  \tr(Y^a Y^{\dagger c}Y^b Y^{\dagger c} ) + 5 \kappa g^2\tr(C_2  Y^aY^{\dagger b} ) + c.c.\nn\\
 \ov{H}^\lambda_{a} =& 2 h_{aef} \tr( m_f Y^{\dagger e} m_f Y^{\dagger f} )+ 4 m^2_{ef}  \tr( Y^a Y^{\dagger e} m_f Y^{\dagger f} ) + c.c.\nn\\
H^Y_{a} =& \tr ( Y_2 (F) m_f^\dagger Y^a m_f^\dagger m_f + Y_2 (F) m_f^\dagger m_f m_f^\dagger Y^a) + c.c.\nn\\
\ov{H}^Y_{a}=& \frac{1}{2} \tr(Y^{\dagger e}  Y^a Y^{\dagger e} m_f m_f^\dagger m_f + 2Y^{\dagger e}  m_f Y^{\dagger e} Y^a  m_f^\dagger m_f + Y^{\dagger e}  m_f Y^{\dagger e} m_f  Y^{\dagger a} m_f ) + c.c.\nn\\
H^3_{a} =&  \tr(Y^a m_f^\dagger  Y^e m_f^\dagger m_f Y^{\dagger e}) + c.c.\nn\\
H^F_{a} =& 4 \tr( C_2 Y^a m_f^\dagger  m_f m_f^\dagger) + c.c.
\end{align}
Here 
\begin{align}
Y_2 (F)_{ij} \equiv& (Y^{\dagger a} Y^a)_{ij}. 
\end{align}

\subsection{Translating from $\ov{\rm MS}$ to $\ov{\rm DR}'$}
\label{SEC:SHIFTS}

To specialise the above expressions to the supersymmetric case, they must be transformed to a complex basis (by summing repeated indices over both raised and lowered indices alternately) and insert the SUSY couplings. These can be written as block diagonal matrices, with the top row/left column corresponding to gauge indices, and bottom row/right column matter indices. The Yukawa matrices become
\begin{align}
Y_i =& \sqrt{2} g\2b2[0,(t^{A})^j_i][(t^{A})^j_i,0] \qquad Y^{\dagger\ i} = \sqrt{2} g\2b2[0,(t^{A})^i_j][(t^{A})^i_j,0] \nn\\
Y^i =& \2b2[0,0][0,Y^{ijk}] \qquad Y^\dagger_i =\2b2[0,0][0,Y_{ijk}]
\end{align}
where $(t^{A})^j_i$ are the gauge generators. With the definitions
\begin{align}
(Y_2)^a_b \equiv& Y^{acd} Y_{bcd}\nn\\
S_2 \delta^{AB} \equiv& (t^{A})^j_i (t^{B})^i_j
\end{align}
then 
\begin{align}
Y_2 (S)^a_b \rightarrow& (Y_2)^a_b +  4 g^2 C_{2\, g}^{\ \ f}\nn\\
Y_2(F) \rightarrow& \2b2[2g^2 S_2\ {\bf 1} ,0][9,2g^2 C_2 + Y_2].
\end{align}

The fermion mass terms then become
\begin{align}
m_f = \2b2[M\ {\bf 1},m_D^{jA}][m_D^{iB},\mu^{ij}].
\end{align}

The one-loop corrections from translating from $\ov{\rm MS}$ to $\ov{\rm DR}'$ only modify the fermionic part. Specialising to the case of interest, where `$a$' is a singlet index, the corrections to the couplings are \cite{Martin:1993yx}:
\begin{align}
\mu \rightarrow& \mu + \frac{g^2}{32\pi^2} [ C_2 (i) + C_2 (j)] \mu_{ij} = \frac{g^2}{32\pi^2} \{ C_2, \mu \}\nn\\
M \rightarrow& M + \frac{g^2}{16\pi^2} C_2 (G) M = \frac{g^2}{32\pi^2} \{ C_2, M \} \nn\\
Y_a^\dagger \rightarrow& Y_a^\dagger \bigg[ 1 + \frac{g^2}{32\pi^2} [ C_2(j) + C_2(k)] \bigg] \nn\\
=& Y_a^\dagger  +  \frac{g^2}{32\pi^2} \{ C_2, Y_a^\dagger \}\nn\\
Y_a \rightarrow& 0 . 
\end{align}
For general Yukawa couplings (not involving a singlet) there are additional contributions; there is also a shift for general quartic couplings. However, they will not be relevant here.

For the Dirac gaugino mass, there is a similar transformation derived via the same technique:
\begin{align}
m_{DiA} \rightarrow& m_{DiA} + \frac{g^2}{16\pi^2} C_2 (G)m_{DiA} = m_{DiA} +\frac{g^2}{32\pi^2} \{ C_2, m_{DiA}\} 
\end{align}
and hence 
\begin{align}
m_f \rightarrow& m_f +  \frac{g^2}{32\pi^2} \{ C_2, m_f\}.
\end{align}
Note that there is no difference for the scalar trilinear couplings  $r^{ij}_{\ \ k}, r_{ij}^{\ \ k} $ between $\ov{\rm MS}$ and $\ov{\rm DR}'$, just as there is none for the couplings $A^{ijk}$.

\subsection{Result}

The expressions can now be transformed to the SUSY basis and the shifts applied as described in subsection \ref{SEC:SHIFTS} to the expressions for the non-susy tadpole RGEs in subsection \ref{SEC:NONSUSY} to obtain the RGEs for the tadpole in the theory described by equations (\ref{EQ:STANDARD}), (\ref{EQ:NONSTANDARD}) and (\ref{EQ:SUPERPOTENTIAL}). After a large amount of tedious algebra, the result for the tadpole beta-functions can be simply written as 
\begin{align}
\beta_{t^a}^{(i)} \equiv& X_S^{(i)} + X_\xi^{(i)} + X_D^{(i)}
\end{align}
where $i$ is the loop order, $X_S^{(i)}$ is the tadpole beta-function involving only standard soft terms, given by
\begin{align}
(4\pi)^2 X^{(1)}_S  =&  (4\pi)^2(\gamma^{(1)})^a_b t^b  \nn\\
&+  A^{aef} ( B_{ef} + Y_{efg} L^g) + Y_{efh} \mu^{ah} B^{ef} + 2Y^{ajk}\mu_{jm}  (m^2)_k^m  
\end{align} 
and
\begin{align}
(4\pi)^4 X^{(2)}_S  =&  (4\pi)^4(\gamma^{(2)})^a_b t^b  \nn\\
&-A^{aef} A_{egh} Y^{ghm} \mu_{mf} -Y^{aem} \mu_{mf}  A_{egh} A^{fgh} - Y_{efm} \mu^{m a} A^{fgh} Y_{ghn} \mu^{ne} \nn\\
&- 4g^2 A^{aef} (C_2)^{m}_{f}  \ov{M} \mu_{me}  -4g^2 Y_{efp} \mu^{ap} (C_2)^{e}_k  \ov{M} \mu^{kf}+8g^2  |M|^2 (C_2)^{j}_{k} \mu_{jl} Y^{akl} \nn\\
&  - 4 g^2 M (B_{ef} + Y_{efp} L^p) (C_2)^{e}_{j} Y^{ajf}\nn\\
&-Y^{aem} (Y_2)^{k}_{m} (m^2)_e^f \mu_{kf}  -2 Y^{ahm} Y_{meg} Y^{fgn} \mu_{nh}  (m^2)_f^e \nn\\
&+ ( 4 g^2 C_2 - Y_2)_g^f \bigg[ Y^{aek} \mu_{kf} (m^2)_e^g + Y^{agk} \mu_{ke} (m^2)_f^e \bigg] \nn\\
&+(4 g^2 C_2 - Y_2)_g^f\bigg[ A^{aeg} (B_{ef} + Y_{efm} L^m) + Y_{efk} \mu^{ka} B^{eg} \bigg] \nn\\
&-Y^{aem} Y_{mgh} (B_{ef} + Y_{efk} L^k) A^{fgh} .
\end{align}
In the above, $(\gamma^{(1)})^a_b $ and $(\gamma^{(2)})^a_b $ are the one- and two-loop chiral superfield anomalous dimensions for singlets respectively, given by
\begin{align}
(4\pi)^2(\gamma^{(1)})^a_b =& \frac{1}{2} (Y_{2})_b^a - 2 g^2 (C_2)_i^j \nn\\
(4\pi)^4(\gamma^{(2)})^a_b =& 2 g^2 (C_2)_k^l Y^{ajk} Y_{bjl} -  \frac{1}{2} Y_{amn}  (Y_{2})_r^n Y^{mrb} . 
\end{align}

The new terms are  
\begin{align}
(4\pi)^2 X^{(1)}_\xi =&2 \sqrt{2} g_Y m_D^{aY} \tr(\mathcal{Y} m^2) \nn\\
 (4\pi)^4 X^{(2)}_\xi =&2 \sqrt{2} g_Y m_D^{aY} \tr( \C{Y} m^2 (4 g^2 C_2 - Y_2))
\end{align}
where $\C{Y}$ is the charge operator of  $U(1)_Y$, and
\begin{align}
(4\pi)^2 X^{(1)}_D =&2 \bigg[(m_D^2)_{ef} (A^{aef}  + M Y^{aef}) + Y_{efk} \mu^{ka} (m_D^2)^{ef} \bigg] \nn\\
(4\pi)^4 X_D^{(2)} =& 4 (\beta_{m_D}^{(1)}/m_D)^f_g \bigg[(m_D^2)_{ef} (A^{aeg}  + M Y^{aeg}) + Y_{efk} \mu^{ka} (m_D^2)^{eg} \bigg] \nn\\
(\beta_{m_D}^{(1)}/m_D)^f_g \equiv& \frac{1}{2} (Y_2)_g^f + g^2 (S_2 - 5C_{2} (G)) \delta^f_g .
\end{align}

The new contributions can be explained as follows. Firstly, the presence of $X_\xi$ is simply due to the renormalisation of the Fayet--Iliopoulos term. Since the FI term is absorbed as it is generated by the running into the soft term and the tadpole (so $\xi =0$), via equation (\ref{EQ:FITADPOLE}), it is found that
\begin{align}
X_\xi =& \frac{d}{d \log \mu} (\sqrt{2} m_D^{a Y} \xi_Y)  = \sqrt{2} m_D^{a Y} \beta_{\xi_Y}.
\end{align}
The result for the RGE of the FI term is then exactly as found in \cite{Jack:2000jr}, \emph{without any additional piece coming from Dirac gaugino terms}. They gave
\begin{align}
16 \pi^2\hat{\beta}_\xi^{(1)} =& 2 g \tr[ \C{Y} m^2] \nn\\
16 \pi^2\hat{\beta}_\xi^{(2)} =& -4 g \tr[ \C{Y} m^2 \gamma^{(1)}] \nn\\
16 \pi^2 (\gamma^{(1)})^j_i =& \frac{1}{2} (Y_{2})_i^j - 2 g^2 (C_2)_i^j 
\end{align}
which clearly exactly agrees with the above. 

Secondly, the $X_D$ terms contain two terms mimicking $B$ insertions -- via equation (\ref{EQ:SHIFTS}) -- in the supergraph when they do not involve a gauge line, but \emph{not} $m^2$ insertions. The additional term proportional to $M (m_D)^2$ is new and could not be obtained from shifting a standard term. Then the two-loop terms involve just wavefunction renormalisation of these. 

Since the tree-level RGE for the tadpole only includes a gauge coupling via the FI term $X_\xi$, the two-loop contributions $X_S^{(2)}, X_D^{(2)}$ include only one gauge coupling, and so the interpretation of the above result for several gauge groups  is straightforward. %

\section{RGEs of a Dirac Gaugino extension of the (N)MSSM}
\label{SEC:MODEL}

Here I present the full two-loop RGEs for a Dirac gaugino extension of the (N)MSSM. For generality, both Dirac and Majorana masses are included, as are all of the phenomenologically interesting superpotential couplings involving the new adjoint superfields, even those that break R-symmetry (this is motivated by generality and the possiblity of generating $\mu$ and $B_\mu$ terms \cite{Benakli:2011kz}). Hence the model encompasses those studied in e.g. \cite{Fox:2002bu}, \cite{Nelson:2002ca} and \cite{Belanger:2009wf,Benakli:2010gi,Benakli:2011kz} (the one-loop RGEs were presented in \cite{Benakli:2010gi} without including Majorana gaugino masses). 

The particle content of the model is just that of the  MSSM  extended by adjoint chiral superfields $\mathbf{S},
\mathbf{T}, \mathbf{O} $
where $S$ is a singlet,  $T = \frac{1}{2}\2b2[T^0,\sqrt{2} T^+][\sqrt{2} T^-,- T^0]$ an $SU(2)$ triplet, and $\mathbf{O}$ a colour octet. The Dirac gaugino masses are denoted $m_{D1}, m_{D2}, m_{D3}$ coupling the respective singlet, triplet and octet fermions to the gauginos of the corresponding group, with gauge couplings $g_1, g_2, g_3$ and Majorana masses $M_1, M_2, M_3$.

The superpotential of the model is:
\begin{eqnarray}
W &=& Y_u \hat{u} \hat{q} H_u - Y_d \hat{d} \hat{q} H_d - Y_e \hat{e} \hat{l} H_d  +  \mu \mathbf{H_u\!\cdot\! H_d }\nn\\
&&  + \lambda_S \mathbf{SH_u\!\cdot\!H_d}  + 2  \lambda_T \mathbf{H_d\!\cdot\! T H_u} \nn\\
&&+L \mathbf{S}  + \frac {M_S}{2}\mathbf{S}^2 + \frac{\kappa}{3}
\mathbf{S}^3 + M_T \textrm{tr}(\mathbf{TT}) + M_O \textrm{tr}(\mathbf{OO}),
\label{NewSuperPotential}
\end{eqnarray}
the usual scalar soft terms are
\begin{align}
\label{potential4}
- \Delta\mathcal{L}^{\rm scalar\ soft}_{\rm MSSM} =& [ A_u \hat{u} \hat{q} H_u - A_d \hat{d} \hat{q} H_d - A_e \hat{e} \hat{l} H_d  + h.c. ]\nn\\
&+ m_{H_u}^2 |H_u|^2 +
m_{H_d}^2 |H_d|^2
 + [B_{\mu} H_u\cdot H_d + h.c. ]\nn\\
&+ \hat{q}^i (m_q^2)_i^j \hat{q}_j + \hat{u}^i (m_u^2)_i^j \hat{u}_j + \hat{d}^i (m_d^2)_i^j \hat{d}_j + \hat{l}^i (m_l^2)_i^j \hat{l}_j  + \hat{e}^i (m_e^2)_i^j \hat{e}_j  
\end{align}
and there are  soft terms involving the adjoint scalars
\begin{eqnarray}
- \Delta\mathcal{L}^{\rm scalar\ soft}_{\rm adjoints} &= &  (t_S S + h.c.) \nn\\
&&+ m_S^2  |S|^2 + \frac{1}{2} B_S
(S^2 + h.c.)  + 2 m_T^2 \textrm{tr}(T^\dagger T) + (B_T \textrm{tr}(T T)+ h.c.)
\nonumber \\  &&+ 
[A_S  \lambda_S SH_u\cdot H_d +  2 A_T  H_d \cdot T H_u +
\frac{1}{3} \kappa  A_{\kappa} S^3 + h.c.] \nonumber \\ &&+ 2 m_O^2 \textrm{tr}(O^\dagger O) 
+ (B_O \textrm{tr}(OO)+ h.c.)
\label{Lsoft-DGAdjoint}
\end{eqnarray}
with the definition $H_u\cdot H_d = H^+_uH^-_d - H^0_u H^0_d$.

The most
general renormalisable Lagrangian would include additional superpotential
interactions\footnote{Note there are no terms
$\textrm{tr}(\mathbf{T}),\textrm{tr}(\mathbf{O}),
\textrm{tr}(\mathbf{TTT})$ since these vanish by gauge invariance.}
\begin{align}
W_{2} =  \lambda_{ST}
\mathbf{S}\textrm{tr}(\mathbf{TT}) +\lambda_{SO} \mathbf{S}\textrm{tr}(\mathbf{OO})
 + \frac{\kappa_O}{3} \textrm{tr}(\mathbf{OOO}).
\label{AdjointSuperpotential}\end{align}
as well as the equivalent adjoint scalar A-terms. It would be straightforward to add these (and indeed  $\lambda_{ST}, \lambda_{SO}$ would contribute to the singlet tadpole) but they violate R-symmetry (admittedly as do all the terms on the last line of equation \ref{NewSuperPotential}) and, more importantly, are much less phenomenologically interesting than the other terms. 

To determine the parameters of the model, the standard soft terms and Dirac gaugino masses can be run according to the RGEs given below, and then at the scale of interest determine the physical soft masses by applying the shifts
\begin{align}
m_S^2 \rightarrow&  m_S^2 + 2 |m_{D1}|^2 \nn\\
B_S^2 \rightarrow&  B_S^2 + 2 m_{D1}^2
\end{align}
and similarly for $m_T, B_T, m_O, B_O$.
 
The standard soft and supersymmetric RGEs, presented in the appendix, were calculated by implementing the model in SARAH  \cite{Staub:2008uz,Staub:2010jh} version 3.0.41. The RGEs for the Dirac mass terms and tadpoles are presented below, using the results of the previous section and the anomalous dimensions of the adjoint superfields.

\subsection{Dirac gaugino masses}

Note $g_Y = \sqrt{\frac{3}{5}} g_1$; then
\begin{align}
(4\pi)^2 \beta_{m_{D1}}^{(1)} =& m_{D1} \bigg[2 \Big(|\kappa_S|^2 + |\lambda_S|^2\Big) + \frac{33}{5} g_{1}^{2} \bigg]\\
(4\pi)^4 \beta_{m_{D1}}^{(2)} =& m_{D1} \bigg[-8 |\kappa_{S}|^{4} -8 |\lambda_S|^2 |\kappa_S|^2  \nonumber \\ 
 &-\frac{2}{5} |\lambda_S|^2 \Big(10 |\lambda_S|^2  -15 g_{2}^{2}  + 15 \mbox{Tr}\Big({Y_d  Y_{d}^{\dagger}}\Big)  + 15 \mbox{Tr}\Big({Y_u  Y_{u}^{\dagger}}\Big)  + 30 |\lambda_T|^2     + 5 \mbox{Tr}\Big({Y_e  Y_{e}^{\dagger}}\Big) \Big)\nn\\ 
&+\frac{1}{25} g_{1}^{2} \Big(-130 \mbox{Tr}\Big({Y_u  Y_{u}^{\dagger}}\Big)  + 135 g_{2}^{2}  + 199 g_{1}^{2}   + 440 g_{3}^{2}  -70 \mbox{Tr}\Big({Y_d  Y_{d}^{\dagger}}\Big) \nn\\
& -90 |\lambda_T|^2  -90 \mbox{Tr}\Big({Y_e  Y_{e}^{\dagger}}\Big) \Big)\bigg]\nn\\ 
(4\pi)^2 \beta_{m_{D2}}^{(1)} =& m_{D2} \bigg[2 |\lambda_T|^2  - g_{2}^{2} \bigg]\nn\\
(4\pi)^4 \beta_{m_{D2}}^{(2)} =& m_{D2} \bigg[28 g_{2}^{4} -12 |\lambda_{T}|^{4}  \nonumber \\ 
 &+\frac{2}{5} |\lambda_T|^2 \Big(-10 |\lambda_S|^2   -15 \mbox{Tr}\Big({Y_d  Y_{d}^{\dagger}}\Big)  -15 \mbox{Tr}\Big({Y_u  Y_{u}^{\dagger}}\Big)  + 3 g_{1}^{2}  -5 g_{2}^{2}  -5 \mbox{Tr}\Big({Y_e  Y_{e}^{\dagger}}\Big) \Big) \nn\\
&+\frac{1}{5} g_{2}^{2} \Big(-10 |\lambda_S|^2  -10 \mbox{Tr}\Big({Y_e  Y_{e}^{\dagger}}\Big)  + 120 g_{3}^{2}  + 245 g_{2}^{2}  -30 \mbox{Tr}\Big({Y_d  Y_{d}^{\dagger}}\Big)  \nn\\
&-30 \mbox{Tr}\Big({Y_u  Y_{u}^{\dagger}}\Big)  -70 |\lambda_T|^2  + 9 g_{1}^{2} \Big)\nn\bigg]\nn\\
(4\pi)^2 \beta_{m_{D3}}^{(1)} =&m_{D3} \bigg[-6 g_{3}^{2} \bigg]\nn\\
(4\pi)^4 \beta_{m_{D3}}^{(2)} =& m_{D3} \bigg[36 g_{3}^{4} + \frac{1}{5} g_{3}^{2} \Big(11 g_{1}^{2}  -20 \mbox{Tr}\Big({Y_d  Y_{d}^{\dagger}}\Big)  -20 \mbox{Tr}\Big({Y_u  Y_{u}^{\dagger}}\Big)  + 340 g_{3}^{2}  + 45 g_{2}^{2} \Big)\bigg]\nn
\end{align}

\subsection{Tadpole equation}

The one-loop contribution to $X_S$ is given by
\begin{align}
(4\pi)^2 X_S^{(1)}  =&  (2|\kappa_S|^2 +2|\lambda_S|^2 )t_S \nn\\
&+4 m_S^2 \kappa_S M_S^* +2M_S B_S \kappa_S^* +4 M_S B_{\mu} \lambda_S^* +4 \lambda_S \mu^* (m_{H_d}^2  + m_{H_u}^2) \nn\\
& +4 L_S \kappa_S^* A_\kappa +2B_S^* A_\kappa +4 L_S \lambda_S^* A_S +4 B_{\mu}^* A_S .
\end{align}
The two-loop contribution is much longer and is given in the appendix, equation (\ref{EQ:STANDARDTADPOLES}).

For the Dirac gaugino contribution arising via the FI term define
\begin{align}
\sigma_{1,1} \equiv& \sqrt{\frac{3}{5}} g_1\tr( \C{Y} m^2 ) \nn\\
\sigma_{3,1} \equiv& \frac{1}{4} \sqrt{\frac{3}{5}} g_1 \tr( \C{Y} m^2 (4 g^2 C_2 - Y_2)).
\end{align}
 $\sigma_{1,1}$ and $\sigma_{3,1}$ also appear in the scalar mass RGEs; the full expressions for these in this model are given in appendix equation (\ref{EQ:SIGMAS}).
Then
\begin{align}
(4\pi)^2 X_\xi^{(1)} 
=& 2 \sqrt{2} m_{D1}\sigma_{1,1} \nn\\
(4\pi)^4X_\xi^{(2)}
=& 8 \sqrt{2} m_{D1}  \sigma_{3,1}.
\end{align}

Finally, the Dirac gaugino parts are given by
\begin{align}
(4\pi)^2 X_D^{(1)}
=& 4  (m_{D1}^*)^2 \bigg[ A_\kappa + M_1 \kappa_S \bigg] + 4 m_{D1}^2 \kappa_S^* M_S  \\
(4\pi)^4 X_D^{(2)}
=& 8 \bigg[2 \Big(|\kappa_S|^2 + |\lambda_S|^2\Big) + \frac{33}{5} g_{1}^{2} \bigg]\bigg[  (m_{D1}^*)^2 \bigg( A_\kappa + M_1 \kappa_S \bigg) + m_{D1}^2 \kappa_S^* M_S \bigg]. \nn
\end{align}

The conclusion is that for R-symmetric soft terms that obey $\tr(\C{Y} m^2) =0$ at some scale (such as, for example, in gauge mediation with R-symmetric F-terms \cite{Amigo:2008rc,Benakli:2008pg}), the new Dirac gaugino mass-dependent contributions to the singlet tadpole will be safely negligible.

\section{Conclusions}

The set of two-loop RGEs for Dirac gaugino models is now complete, which opens up the possibility of implementing them in spectrum generators (such as by an extension of \cite{Staub:2008uz,Staub:2010jh}) and studying their precision phenomenology. Since Dirac mass terms only appear in supersoft operators (in spontaneously broken supersymmetric theories) their effect upon the RGEs is extremely mild. In particular, they remain on an RG invariant trajectory with respect to the mass squared and $B$ terms, but the result of this paper is that, although they contribute new terms to the tadpole RGE, these are exhausted at one-loop and the two-loop correction is merely a particular form of wavefunction renormalisation of the one-loop term. 

As mentioned in the introduction, the tadpole RGE is particularly important for studying the Higgs potential, and since the new Higgs couplings in Dirac gaugino models can allow a natural increase in the Higgs mass -- alleviating fine-tuning -- it is now imperative to study the Higgs sector of Dirac gaugino models including full loop corrections \cite{INPROG}.

\section*{Acknowledgements}

I was supported by ERC advanced grant 226371. I thank I.~Jack and D.~R.~T.~Jones for helpful discussions. I would also like to thank the Isaac Newton Institute, and the organisers of the string phenomenology workshop in the ``Mathematics and Applications of Branes in String and M-theory'' programme, for excellent hospitality and providing a stimulating environment while this project was completed. Finally I thank K.~A.~West for careful reading of the manuscript.

\appendix

\section{Two-loop RGEs for SUSY and standard soft terms}

In this appendix the RGEs for the standard soft terms in the model of section \ref{SEC:MODEL} are presented. For brevity, the factors of $(4\pi)^2$ and $(4\pi)^4$ for one- and two-loop quantities shall be dropped. All of the below were generated using SARAH \cite{Staub:2008uz,Staub:2010jh} version 3.0.41 and so are presented in the notation of that package.

\subsection{Anomalous Dimensions}
{\allowdisplaybreaks \begin{align} 
\gamma_{\hat{q}}^{(1)} & =  
-\frac{1}{30} \Big(45 g_{2}^{2}  + 80 g_{3}^{2}  + g_{1}^{2}\Big){\bf 1}  + {Y_{d}^{\dagger}  Y_d} + {Y_{u}^{\dagger}  Y_u}\\ 
\gamma_{\hat{q}}^{(2)} & =  
+\Big(8 g_{2}^{2} g_{3}^{2}  + \frac{1}{90} g_{1}^{2} \Big(16 g_{3}^{2}  + 9 g_{2}^{2} \Big) + \frac{199}{900} g_{1}^{4}  + \frac{27}{4} g_{2}^{4}  + \frac{64}{9} g_{3}^{4} \Big){\bf 1} +\frac{4}{5} g_{1}^{2} {Y_{u}^{\dagger}  Y_u} - |\lambda_S|^2 {Y_{u}^{\dagger}  Y_u} \nonumber \\ 
 &-3 |\lambda_T|^2 {Y_{u}^{\dagger}  Y_u} -2 {Y_{d}^{\dagger}  Y_d  Y_{d}^{\dagger}  Y_d} -2 {Y_{u}^{\dagger}  Y_u  Y_{u}^{\dagger}  Y_u} \nonumber \\ 
 &+{Y_{d}^{\dagger}  Y_d} \Big(-3 |\lambda_T|^2  -3 \mbox{Tr}\Big({Y_d  Y_{d}^{\dagger}}\Big)  + \frac{2}{5} g_{1}^{2}  - |\lambda_S|^2  - \mbox{Tr}\Big({Y_e  Y_{e}^{\dagger}}\Big) \Big)-3 {Y_{u}^{\dagger}  Y_u} \mbox{Tr}\Big({Y_u  Y_{u}^{\dagger}}\Big) \\ 
\gamma_{\hat{l}}^{(1)} & =  
-\frac{3}{10} \Big(5 g_{2}^{2}  + g_{1}^{2}\Big){\bf 1}  + {Y_{e}^{\dagger}  Y_e}\\ 
\gamma_{\hat{l}}^{(2)} & =  
+\frac{9}{100} \Big(10 g_{1}^{2} g_{2}^{2}  + 23 g_{1}^{4}  + 75 g_{2}^{4} \Big){\bf 1} -2 {Y_{e}^{\dagger}  Y_e  Y_{e}^{\dagger}  Y_e} \nonumber \\ 
 &+{Y_{e}^{\dagger}  Y_e} \Big(-3 |\lambda_T|^2  -3 \mbox{Tr}\Big({Y_d  Y_{d}^{\dagger}}\Big)  + \frac{6}{5} g_{1}^{2}  - |\lambda_S|^2  - \mbox{Tr}\Big({Y_e  Y_{e}^{\dagger}}\Big) \Big)\\ 
\gamma_{\hat{H}_d}^{(1)} & =  
3 |\lambda_T|^2  + 3 \mbox{Tr}\Big({Y_d  Y_{d}^{\dagger}}\Big)  -\frac{3}{10} g_{1}^{2}  -\frac{3}{2} g_{2}^{2}  + |\lambda_S|^2 + \mbox{Tr}\Big({Y_e  Y_{e}^{\dagger}}\Big)\\ 
\gamma_{\hat{H}_d}^{(2)} & =  
+\frac{207}{100} g_{1}^{4} +\frac{9}{10} g_{1}^{2} g_{2}^{2} +\frac{27}{4} g_{2}^{4} +12 g_{2}^{2} |\lambda_T|^2 -2 \lambda_S |\kappa_S|^2 \lambda_S^* -3 \lambda_{S}^{2} \lambda_{S}^{*,2} -15 \lambda_{T}^{2} \lambda_{T}^{*,2} \nonumber \\ 
 &-\frac{2}{5} g_{1}^{2} \mbox{Tr}\Big({Y_d  Y_{d}^{\dagger}}\Big) +16 g_{3}^{2} \mbox{Tr}\Big({Y_d  Y_{d}^{\dagger}}\Big) +\frac{6}{5} g_{1}^{2} \mbox{Tr}\Big({Y_e  Y_{e}^{\dagger}}\Big) -9 |\lambda_T|^2 \mbox{Tr}\Big({Y_u  Y_{u}^{\dagger}}\Big) \nonumber \\ 
 &-3 |\lambda_S|^2 \Big(2 \lambda_T \lambda_T^*  + \mbox{Tr}\Big({Y_u  Y_{u}^{\dagger}}\Big)\Big)-9 \mbox{Tr}\Big({Y_d  Y_{d}^{\dagger}  Y_d  Y_{d}^{\dagger}}\Big) -3 \mbox{Tr}\Big({Y_d  Y_{u}^{\dagger}  Y_u  Y_{d}^{\dagger}}\Big) -3 \mbox{Tr}\Big({Y_e  Y_{e}^{\dagger}  Y_e  Y_{e}^{\dagger}}\Big) \\ 
\gamma_{\hat{H}_u}^{(1)} & =  
3 |\lambda_T|^2  -\frac{3}{10} \Big(-10 \mbox{Tr}\Big({Y_u  Y_{u}^{\dagger}}\Big)  + 5 g_{2}^{2}  + g_{1}^{2}\Big) + |\lambda_S|^2\\ 
\gamma_{\hat{H}_u}^{(2)} & =  
+\frac{207}{100} g_{1}^{4} +\frac{9}{10} g_{1}^{2} g_{2}^{2} +\frac{27}{4} g_{2}^{4} +12 g_{2}^{2} |\lambda_T|^2 -2 \lambda_S |\kappa_S|^2 \lambda_S^* -3 \lambda_{S}^{2} \lambda_{S}^{*,2} -15 \lambda_{T}^{2} \lambda_{T}^{*,2} \nonumber \\ 
 &-9 |\lambda_T|^2 \mbox{Tr}\Big({Y_d  Y_{d}^{\dagger}}\Big) -3 |\lambda_T|^2 \mbox{Tr}\Big({Y_e  Y_{e}^{\dagger}}\Big) - |\lambda_S|^2 \Big(3 \mbox{Tr}\Big({Y_d  Y_{d}^{\dagger}}\Big)  + 6 \lambda_T \lambda_T^*  + \mbox{Tr}\Big({Y_e  Y_{e}^{\dagger}}\Big)\Big)\nonumber \\ 
 &+\frac{4}{5} g_{1}^{2} \mbox{Tr}\Big({Y_u  Y_{u}^{\dagger}}\Big) +16 g_{3}^{2} \mbox{Tr}\Big({Y_u  Y_{u}^{\dagger}}\Big) -3 \mbox{Tr}\Big({Y_d  Y_{u}^{\dagger}  Y_u  Y_{d}^{\dagger}}\Big) -9 \mbox{Tr}\Big({Y_u  Y_{u}^{\dagger}  Y_u  Y_{u}^{\dagger}}\Big) \\ 
\gamma_{\hat{d}}^{(1)} & =  
2 {Y_d^*  Y_{d}^{T}}  -\frac{2}{15} \Big(20 g_{3}^{2}  + g_{1}^{2}\Big){\bf 1} \\ 
\gamma_{\hat{d}}^{(2)} & =  
+\frac{2}{225} \Big(101 g_{1}^{4}  + 800 g_{3}^{4}  + 80 g_{1}^{2} g_{3}^{2} \Big){\bf 1} -2 \Big({Y_d^*  Y_{d}^{T}  Y_d^*  Y_{d}^{T}} + {Y_d^*  Y_{u}^{T}  Y_u^*  Y_{d}^{T}}\Big)\nonumber \\ 
 &+{Y_d^*  Y_{d}^{T}} \Big(-2 |\lambda_S|^2  -2 \mbox{Tr}\Big({Y_e  Y_{e}^{\dagger}}\Big)  + 6 g_{2}^{2}  -6 |\lambda_T|^2  -6 \mbox{Tr}\Big({Y_d  Y_{d}^{\dagger}}\Big)  + \frac{2}{5} g_{1}^{2} \Big)\\ 
\gamma_{\hat{u}}^{(1)} & =  
2 {Y_u^*  Y_{u}^{T}}  -\frac{8}{15} \Big(5 g_{3}^{2}  + g_{1}^{2}\Big){\bf 1} \\ 
\gamma_{\hat{u}}^{(2)} & =  
+\frac{8}{225} \Big(107 g_{1}^{4}  + 200 g_{3}^{4}  + 80 g_{1}^{2} g_{3}^{2} \Big){\bf 1} \nonumber \\ 
 &-\frac{2}{5} \Big(5 \Big({Y_u^*  Y_{d}^{T}  Y_d^*  Y_{u}^{T}} + {Y_u^*  Y_{u}^{T}  Y_u^*  Y_{u}^{T}}\Big) + {Y_u^*  Y_{u}^{T}} \Big(-15 g_{2}^{2}  + 15 |\lambda_T|^2  + 15 \mbox{Tr}\Big({Y_u  Y_{u}^{\dagger}}\Big)  + 5 |\lambda_S|^2  + g_{1}^{2}\Big)\Big)\\ 
\gamma_{\hat{e}}^{(1)} & =  
2 {Y_e^*  Y_{e}^{T}}  -\frac{6}{5} g_{1}^{2} {\bf 1} \\ 
\gamma_{\hat{e}}^{(2)} & =  
+\frac{234}{25} g_{1}^{4} {\bf 1} -2 {Y_e^*  Y_{e}^{T}  Y_e^*  Y_{e}^{T}} \nonumber \\ 
 &+{Y_e^*  Y_{e}^{T}} \Big(-2 |\lambda_S|^2  -2 \mbox{Tr}\Big({Y_e  Y_{e}^{\dagger}}\Big)  + 6 g_{2}^{2}  -6 |\lambda_T|^2  -6 \mbox{Tr}\Big({Y_d  Y_{d}^{\dagger}}\Big)  -\frac{6}{5} g_{1}^{2} \Big)\\ 
\gamma_{S}^{(1)} & =  
2 \Big(|\kappa_S|^2 + |\lambda_S|^2\Big)\\ 
\gamma_{S}^{(2)} & =  
-8 \kappa_{S}^{2} \kappa_{S}^{*,2} -8 \lambda_S |\kappa_S|^2 \lambda_S^* \nonumber \\ 
 &-\frac{2}{5} |\lambda_S|^2 \Big(10 \lambda_S \lambda_S^*  -15 g_{2}^{2}  + 15 \mbox{Tr}\Big({Y_d  Y_{d}^{\dagger}}\Big)  + 15 \mbox{Tr}\Big({Y_u  Y_{u}^{\dagger}}\Big)  + 30 \lambda_T \lambda_T^*  -3 g_{1}^{2}  + 5 \mbox{Tr}\Big({Y_e  Y_{e}^{\dagger}}\Big) \Big)\\ 
\gamma_{T}^{(1)} & =  
2 |\lambda_T|^2  -4 g_{2}^{2} \\ 
\gamma_{T}^{(2)} & =  
+28 g_{2}^{4} -12 \lambda_{T}^{2} \lambda_{T}^{*,2} \nonumber \\ 
 &+\frac{2}{5} |\lambda_T|^2 \Big(-10 \lambda_S \lambda_S^*  -15 \mbox{Tr}\Big({Y_d  Y_{d}^{\dagger}}\Big)  -15 \mbox{Tr}\Big({Y_u  Y_{u}^{\dagger}}\Big)  + 3 g_{1}^{2}  -5 g_{2}^{2}  -5 \mbox{Tr}\Big({Y_e  Y_{e}^{\dagger}}\Big) \Big)\\ 
\gamma_{O}^{(1)} & =  
-6 g_{3}^{2} \\ 
\gamma_{O}^{(2)} & =  
36 g_{3}^{4} 
\end{align} } 
\subsection{Gauge Couplings}
{\allowdisplaybreaks  \begin{align} 
\beta_{g_1}^{(1)} & =  
\frac{33}{5} g_{1}^{3} \\ 
\beta_{g_1}^{(2)} & =  
\frac{1}{25} g_{1}^{3} \Big(-130 \mbox{Tr}\Big({Y_u  Y_{u}^{\dagger}}\Big)  + 135 g_{2}^{2}  + 199 g_{1}^{2}  -30 |\lambda_S|^2  + 440 g_{3}^{2}  -70 \mbox{Tr}\Big({Y_d  Y_{d}^{\dagger}}\Big) \nn\\
& -90 |\lambda_T|^2  -90 \mbox{Tr}\Big({Y_e  Y_{e}^{\dagger}}\Big) \Big)\\ 
\beta_{g_2}^{(1)} & =  
3 g_{2}^{3} \\ 
\beta_{g_2}^{(2)} & =  
\frac{1}{5} g_{2}^{3} \Big(-10 |\lambda_S|^2  -10 \mbox{Tr}\Big({Y_e  Y_{e}^{\dagger}}\Big)  + 120 g_{3}^{2}  + 245 g_{2}^{2}  -30 \mbox{Tr}\Big({Y_d  Y_{d}^{\dagger}}\Big) \nn\\
& -30 \mbox{Tr}\Big({Y_u  Y_{u}^{\dagger}}\Big)  -70 |\lambda_T|^2  + 9 g_{1}^{2} \Big)\\ 
\beta_{g_3}^{(1)} & =  
0\\ 
\beta_{g_3}^{(2)} & =  
\frac{1}{5} g_{3}^{3} \Big(11 g_{1}^{2}  -20 \mbox{Tr}\Big({Y_d  Y_{d}^{\dagger}}\Big)  -20 \mbox{Tr}\Big({Y_u  Y_{u}^{\dagger}}\Big)  + 340 g_{3}^{2}  + 45 g_{2}^{2} \Big)
\end{align}} 
\subsection{Gaugino Mass Parameters}
{\allowdisplaybreaks  \begin{align} 
\beta_{M_1}^{(1)} & =  
\frac{66}{5} g_{1}^{2} M_1 \\ 
\beta_{M_1}^{(2)} & =  
\frac{2}{25} g_{1}^{2} \Big(398 g_{1}^{2} M_1 +135 g_{2}^{2} M_1 +440 g_{3}^{2} M_1 +440 g_{3}^{2} M_3 +135 g_{2}^{2} M_2 \nn\\
&-30 \lambda_S^* \Big(- A_S  + M_1 \lambda_S \Big)-90 \lambda_T^* \Big(- A_T  + M_1 \lambda_T \Big)\nonumber \\ 
 &-70 M_1 \mbox{Tr}\Big({Y_d  Y_{d}^{\dagger}}\Big) -90 M_1 \mbox{Tr}\Big({Y_e  Y_{e}^{\dagger}}\Big) -130 M_1 \mbox{Tr}\Big({Y_u  Y_{u}^{\dagger}}\Big) +70 \mbox{Tr}\Big({Y_{d}^{\dagger}  A_d}\Big) +90 \mbox{Tr}\Big({Y_{e}^{\dagger}  A_e}\Big) \nonumber \\ 
 &+130 \mbox{Tr}\Big({Y_{u}^{\dagger}  A_u}\Big) \Big)\\ 
\beta_{M_2}^{(1)} & =  
6 g_{2}^{2} M_2 \\ 
\beta_{M_2}^{(2)} & =  
\frac{2}{5} g_{2}^{2} \Big(9 g_{1}^{2} M_1 +120 g_{3}^{2} M_3 +9 g_{1}^{2} M_2 +490 g_{2}^{2} M_2 +120 g_{3}^{2} M_2\nn\\
& -10 \lambda_S^* \Big(- A_S  + M_2 \lambda_S \Big)-70 \lambda_T^* \Big(- A_T  + M_2 \lambda_T \Big)\nonumber \\ 
 &-30 M_2 \mbox{Tr}\Big({Y_d  Y_{d}^{\dagger}}\Big) -10 M_2 \mbox{Tr}\Big({Y_e  Y_{e}^{\dagger}}\Big) -30 M_2 \mbox{Tr}\Big({Y_u  Y_{u}^{\dagger}}\Big) +30 \mbox{Tr}\Big({Y_{d}^{\dagger}  A_d}\Big) +10 \mbox{Tr}\Big({Y_{e}^{\dagger}  A_e}\Big) \nonumber \\ 
 &+30 \mbox{Tr}\Big({Y_{u}^{\dagger}  A_u}\Big) \Big)\\ 
\beta_{M_3}^{(1)} & =  
0\\ 
\beta_{M_3}^{(2)} & =  
\frac{2}{5} g_{3}^{2} \Big(11 g_{1}^{2} M_1 +11 g_{1}^{2} M_3 +45 g_{2}^{2} M_3 +680 g_{3}^{2} M_3 +45 g_{2}^{2} M_2 -20 M_3 \mbox{Tr}\Big({Y_d  Y_{d}^{\dagger}}\Big) \nn\\
&-20 M_3 \mbox{Tr}\Big({Y_u  Y_{u}^{\dagger}}\Big) +20 \mbox{Tr}\Big({Y_{d}^{\dagger}  A_d}\Big) +20 \mbox{Tr}\Big({Y_{u}^{\dagger}  A_u}\Big) \Big)
\end{align}} 
\subsection{Trilinear Superpotential Parameters}
{\allowdisplaybreaks  \begin{align} 
\beta_{Y_u}^{(1)} & =  
3 {Y_u  Y_{u}^{\dagger}  Y_u}  + Y_u \Big(-3 g_{2}^{2}  + 3 |\lambda_T|^2  + 3 \mbox{Tr}\Big({Y_u  Y_{u}^{\dagger}}\Big)  -\frac{13}{15} g_{1}^{2}  -\frac{16}{3} g_{3}^{2}  + |\lambda_S|^2\Big) + {Y_u  Y_{d}^{\dagger}  Y_d}\\ 
\beta_{Y_u}^{(2)} & =  
+\frac{2}{5} g_{1}^{2} {Y_u  Y_{u}^{\dagger}  Y_u} +6 g_{2}^{2} {Y_u  Y_{u}^{\dagger}  Y_u} -3 |\lambda_S|^2 {Y_u  Y_{u}^{\dagger}  Y_u} -9 |\lambda_T|^2 {Y_u  Y_{u}^{\dagger}  Y_u} \nonumber \\ 
 &-2 {Y_u  Y_{d}^{\dagger}  Y_d  Y_{d}^{\dagger}  Y_d} -2 {Y_u  Y_{d}^{\dagger}  Y_d  Y_{u}^{\dagger}  Y_u} -4 {Y_u  Y_{u}^{\dagger}  Y_u  Y_{u}^{\dagger}  Y_u} \nonumber \\ 
 &+{Y_u  Y_{d}^{\dagger}  Y_d} \Big(-3 |\lambda_T|^2  -3 \mbox{Tr}\Big({Y_d  Y_{d}^{\dagger}}\Big)  + \frac{2}{5} g_{1}^{2}  - |\lambda_S|^2  - \mbox{Tr}\Big({Y_e  Y_{e}^{\dagger}}\Big) \Big)\nonumber \\ 
 &-9 {Y_u  Y_{u}^{\dagger}  Y_u} \mbox{Tr}\Big({Y_u  Y_{u}^{\dagger}}\Big) \nonumber \\ 
 &+Y_u \Big(\frac{2743}{450} g_{1}^{4} +g_{1}^{2} g_{2}^{2} +\frac{27}{2} g_{2}^{4} +\frac{136}{45} g_{1}^{2} g_{3}^{2} +8 g_{2}^{2} g_{3}^{2} +\frac{128}{9} g_{3}^{4} +12 g_{2}^{2} |\lambda_T|^2 -2 \lambda_S |\kappa_S|^2 \lambda_S^* \nonumber \\ 
 &-3 \lambda_{S}^{2} \lambda_{S}^{*,2} -15 \lambda_{T}^{2} \lambda_{T}^{*,2} -9 |\lambda_T|^2 \mbox{Tr}\Big({Y_d  Y_{d}^{\dagger}}\Big) -3 |\lambda_T|^2 \mbox{Tr}\Big({Y_e  Y_{e}^{\dagger}}\Big) \nonumber \\ 
 &- |\lambda_S|^2 \Big(3 \mbox{Tr}\Big({Y_d  Y_{d}^{\dagger}}\Big)  + 6 \lambda_T \lambda_T^*  + \mbox{Tr}\Big({Y_e  Y_{e}^{\dagger}}\Big)\Big)+\frac{4}{5} g_{1}^{2} \mbox{Tr}\Big({Y_u  Y_{u}^{\dagger}}\Big) +16 g_{3}^{2} \mbox{Tr}\Big({Y_u  Y_{u}^{\dagger}}\Big) \nonumber \\ 
 &-3 \mbox{Tr}\Big({Y_d  Y_{u}^{\dagger}  Y_u  Y_{d}^{\dagger}}\Big) -9 \mbox{Tr}\Big({Y_u  Y_{u}^{\dagger}  Y_u  Y_{u}^{\dagger}}\Big) \Big)\\ 
\beta_{Y_d}^{(1)} & =  
3 {Y_d  Y_{d}^{\dagger}  Y_d}  + Y_d \Big(-3 g_{2}^{2}  + 3 |\lambda_T|^2  + 3 \mbox{Tr}\Big({Y_d  Y_{d}^{\dagger}}\Big)  -\frac{16}{3} g_{3}^{2}  -\frac{7}{15} g_{1}^{2}  + |\lambda_S|^2 + \mbox{Tr}\Big({Y_e  Y_{e}^{\dagger}}\Big)\Big) + {Y_d  Y_{u}^{\dagger}  Y_u}\\ 
\beta_{Y_d}^{(2)} & =  
+\frac{4}{5} g_{1}^{2} {Y_d  Y_{u}^{\dagger}  Y_u} - |\lambda_S|^2 {Y_d  Y_{u}^{\dagger}  Y_u} -3 |\lambda_T|^2 {Y_d  Y_{u}^{\dagger}  Y_u} -4 {Y_d  Y_{d}^{\dagger}  Y_d  Y_{d}^{\dagger}  Y_d} \nonumber \\ 
 &-2 {Y_d  Y_{u}^{\dagger}  Y_u  Y_{d}^{\dagger}  Y_d} -2 {Y_d  Y_{u}^{\dagger}  Y_u  Y_{u}^{\dagger}  Y_u} \nonumber \\ 
 &+{Y_d  Y_{d}^{\dagger}  Y_d} \Big(-3 |\lambda_S|^2  -3 \mbox{Tr}\Big({Y_e  Y_{e}^{\dagger}}\Big)  + 6 g_{2}^{2}  -9 |\lambda_T|^2  -9 \mbox{Tr}\Big({Y_d  Y_{d}^{\dagger}}\Big)  + \frac{4}{5} g_{1}^{2} \Big)\nonumber \\ 
 &-3 {Y_d  Y_{u}^{\dagger}  Y_u} \mbox{Tr}\Big({Y_u  Y_{u}^{\dagger}}\Big) \nonumber \\ 
 &+Y_d \Big(\frac{287}{90} g_{1}^{4} +g_{1}^{2} g_{2}^{2} +\frac{27}{2} g_{2}^{4} +\frac{8}{9} g_{1}^{2} g_{3}^{2} +8 g_{2}^{2} g_{3}^{2} +\frac{128}{9} g_{3}^{4} +12 g_{2}^{2} |\lambda_T|^2 -2 \lambda_S |\kappa_S|^2 \lambda_S^* \nonumber \\ 
 &-3 \lambda_{S}^{2} \lambda_{S}^{*,2} -15 \lambda_{T}^{2} \lambda_{T}^{*,2} -\frac{2}{5} g_{1}^{2} \mbox{Tr}\Big({Y_d  Y_{d}^{\dagger}}\Big) +16 g_{3}^{2} \mbox{Tr}\Big({Y_d  Y_{d}^{\dagger}}\Big) +\frac{6}{5} g_{1}^{2} \mbox{Tr}\Big({Y_e  Y_{e}^{\dagger}}\Big) \nonumber \\ 
 &-9 |\lambda_T|^2 \mbox{Tr}\Big({Y_u  Y_{u}^{\dagger}}\Big) -3 |\lambda_S|^2 \Big(2 \lambda_T \lambda_T^*  + \mbox{Tr}\Big({Y_u  Y_{u}^{\dagger}}\Big)\Big)-9 \mbox{Tr}\Big({Y_d  Y_{d}^{\dagger}  Y_d  Y_{d}^{\dagger}}\Big) -3 \mbox{Tr}\Big({Y_d  Y_{u}^{\dagger}  Y_u  Y_{d}^{\dagger}}\Big) \nonumber \\ 
 &-3 \mbox{Tr}\Big({Y_e  Y_{e}^{\dagger}  Y_e  Y_{e}^{\dagger}}\Big) \Big)\\ 
\beta_{Y_e}^{(1)} & =  
3 {Y_e  Y_{e}^{\dagger}  Y_e}  + Y_e \Big(-3 g_{2}^{2}  + 3 |\lambda_T|^2  + 3 \mbox{Tr}\Big({Y_d  Y_{d}^{\dagger}}\Big)  -\frac{9}{5} g_{1}^{2}  + |\lambda_S|^2 + \mbox{Tr}\Big({Y_e  Y_{e}^{\dagger}}\Big)\Big)\\ 
\beta_{Y_e}^{(2)} & =  
-4 {Y_e  Y_{e}^{\dagger}  Y_e  Y_{e}^{\dagger}  Y_e} +{Y_e  Y_{e}^{\dagger}  Y_e} \Big(-3 |\lambda_S|^2  -3 \mbox{Tr}\Big({Y_e  Y_{e}^{\dagger}}\Big)  + 6 g_{2}^{2}  -9 |\lambda_T|^2  -9 \mbox{Tr}\Big({Y_d  Y_{d}^{\dagger}}\Big) \Big)\nonumber \\ 
 &+Y_e \Big(\frac{27}{2} g_{1}^{4} +\frac{9}{5} g_{1}^{2} g_{2}^{2} +\frac{27}{2} g_{2}^{4} +12 g_{2}^{2} |\lambda_T|^2 -2 \lambda_S |\kappa_S|^2 \lambda_S^* -3 \lambda_{S}^{2} \lambda_{S}^{*,2} -15 \lambda_{T}^{2} \lambda_{T}^{*,2} \nonumber \\ 
 &-\frac{2}{5} g_{1}^{2} \mbox{Tr}\Big({Y_d  Y_{d}^{\dagger}}\Big) +16 g_{3}^{2} \mbox{Tr}\Big({Y_d  Y_{d}^{\dagger}}\Big) +\frac{6}{5} g_{1}^{2} \mbox{Tr}\Big({Y_e  Y_{e}^{\dagger}}\Big) -9 |\lambda_T|^2 \mbox{Tr}\Big({Y_u  Y_{u}^{\dagger}}\Big) \nonumber \\ 
 &-3 |\lambda_S|^2 \Big(2 \lambda_T \lambda_T^*  + \mbox{Tr}\Big({Y_u  Y_{u}^{\dagger}}\Big)\Big)-9 \mbox{Tr}\Big({Y_d  Y_{d}^{\dagger}  Y_d  Y_{d}^{\dagger}}\Big) -3 \mbox{Tr}\Big({Y_d  Y_{u}^{\dagger}  Y_u  Y_{d}^{\dagger}}\Big) -3 \mbox{Tr}\Big({Y_e  Y_{e}^{\dagger}  Y_e  Y_{e}^{\dagger}}\Big) \Big)\\ 
\beta_{\lambda_T}^{(1)} & =  
2 \lambda_T |\lambda_S|^2  + 3 \lambda_T \mbox{Tr}\Big({Y_d  Y_{d}^{\dagger}}\Big)  + 3 \lambda_T \mbox{Tr}\Big({Y_u  Y_{u}^{\dagger}}\Big)  -7 g_{2}^{2} \lambda_T  + 8 \lambda_{T}^{2} \lambda_T^*  -\frac{3}{5} g_{1}^{2} \lambda_T  + \lambda_T \mbox{Tr}\Big({Y_e  Y_{e}^{\dagger}}\Big) \\ 
\beta_{\lambda_T}^{(2)} & =  
-\frac{1}{50} \lambda_T \Big(-207 g_{1}^{4} -90 g_{1}^{2} g_{2}^{2} -2075 g_{2}^{4} -60 g_{1}^{2} |\lambda_T|^2 -1100 g_{2}^{2} |\lambda_T|^2 +200 \lambda_S |\kappa_S|^2 \lambda_S^* +300 \lambda_{S}^{2} \lambda_{S}^{*,2} \nonumber \\ 
 &+2100 \lambda_{T}^{2} \lambda_{T}^{*,2} +20 g_{1}^{2} \mbox{Tr}\Big({Y_d  Y_{d}^{\dagger}}\Big) -800 g_{3}^{2} \mbox{Tr}\Big({Y_d  Y_{d}^{\dagger}}\Big) +750 |\lambda_T|^2 \mbox{Tr}\Big({Y_d  Y_{d}^{\dagger}}\Big) \nonumber \\ 
 &-60 g_{1}^{2} \mbox{Tr}\Big({Y_e  Y_{e}^{\dagger}}\Big) +250 |\lambda_T|^2 \mbox{Tr}\Big({Y_e  Y_{e}^{\dagger}}\Big) -40 g_{1}^{2} \mbox{Tr}\Big({Y_u  Y_{u}^{\dagger}}\Big) -800 g_{3}^{2} \mbox{Tr}\Big({Y_u  Y_{u}^{\dagger}}\Big) \nonumber \\ 
 &+750 |\lambda_T|^2 \mbox{Tr}\Big({Y_u  Y_{u}^{\dagger}}\Big) +50 |\lambda_S|^2 \Big(16 \lambda_T \lambda_T^*  + 3 \mbox{Tr}\Big({Y_d  Y_{d}^{\dagger}}\Big)  + 3 \mbox{Tr}\Big({Y_u  Y_{u}^{\dagger}}\Big)  + \mbox{Tr}\Big({Y_e  Y_{e}^{\dagger}}\Big)\Big)\nonumber \\ 
 &+450 \mbox{Tr}\Big({Y_d  Y_{d}^{\dagger}  Y_d  Y_{d}^{\dagger}}\Big) +300 \mbox{Tr}\Big({Y_d  Y_{u}^{\dagger}  Y_u  Y_{d}^{\dagger}}\Big) +150 \mbox{Tr}\Big({Y_e  Y_{e}^{\dagger}  Y_e  Y_{e}^{\dagger}}\Big) +450 \mbox{Tr}\Big({Y_u  Y_{u}^{\dagger}  Y_u  Y_{u}^{\dagger}}\Big) \Big)\\ 
\beta_{\lambda_S}^{(1)} & =  
-\frac{3}{5} g_{1}^{2} \lambda_S -3 g_{2}^{2} \lambda_S +2 \lambda_S |\kappa_S|^2 +6 \lambda_S |\lambda_T|^2 +4 \lambda_{S}^{2} \lambda_S^* +3 \lambda_S \mbox{Tr}\Big({Y_d  Y_{d}^{\dagger}}\Big) +\lambda_S \mbox{Tr}\Big({Y_e  Y_{e}^{\dagger}}\Big) \nonumber \\ 
 &+3 \lambda_S \mbox{Tr}\Big({Y_u  Y_{u}^{\dagger}}\Big) \\ 
\beta_{\lambda_S}^{(2)} & =  
-\frac{1}{50} \lambda_S \Big(-207 g_{1}^{4} -90 g_{1}^{2} g_{2}^{2} -675 g_{2}^{4} -1200 g_{2}^{2} |\lambda_T|^2 +400 \kappa_{S}^{2} \kappa_{S}^{*,2} +600 \lambda_S |\kappa_S|^2 \lambda_S^* +500 \lambda_{S}^{2} \lambda_{S}^{*,2} \nonumber \\ 
 &+1500 \lambda_{T}^{2} \lambda_{T}^{*,2} +20 g_{1}^{2} \mbox{Tr}\Big({Y_d  Y_{d}^{\dagger}}\Big) -800 g_{3}^{2} \mbox{Tr}\Big({Y_d  Y_{d}^{\dagger}}\Big) +450 |\lambda_T|^2 \mbox{Tr}\Big({Y_d  Y_{d}^{\dagger}}\Big) \nonumber \\ 
 &-60 g_{1}^{2} \mbox{Tr}\Big({Y_e  Y_{e}^{\dagger}}\Big) +150 |\lambda_T|^2 \mbox{Tr}\Big({Y_e  Y_{e}^{\dagger}}\Big) \nonumber \\ 
 &-30 |\lambda_S|^2 \Big(10 g_{2}^{2}  -15 \mbox{Tr}\Big({Y_d  Y_{d}^{\dagger}}\Big)  -15 \mbox{Tr}\Big({Y_u  Y_{u}^{\dagger}}\Big)  + 2 g_{1}^{2}  -40 \lambda_T \lambda_T^*  -5 \mbox{Tr}\Big({Y_e  Y_{e}^{\dagger}}\Big) \Big)\nonumber \\ 
 &-40 g_{1}^{2} \mbox{Tr}\Big({Y_u  Y_{u}^{\dagger}}\Big) -800 g_{3}^{2} \mbox{Tr}\Big({Y_u  Y_{u}^{\dagger}}\Big) +450 |\lambda_T|^2 \mbox{Tr}\Big({Y_u  Y_{u}^{\dagger}}\Big) +450 \mbox{Tr}\Big({Y_d  Y_{d}^{\dagger}  Y_d  Y_{d}^{\dagger}}\Big) \nonumber \\ 
 &+300 \mbox{Tr}\Big({Y_d  Y_{u}^{\dagger}  Y_u  Y_{d}^{\dagger}}\Big) +150 \mbox{Tr}\Big({Y_e  Y_{e}^{\dagger}  Y_e  Y_{e}^{\dagger}}\Big) +450 \mbox{Tr}\Big({Y_u  Y_{u}^{\dagger}  Y_u  Y_{u}^{\dagger}}\Big) \Big)\\ 
\beta_{\kappa_S}^{(1)} & =  
6 \kappa_S \Big(|\kappa_S|^2 + |\lambda_S|^2\Big)\\ 
\beta_{\kappa_S}^{(2)} & =  
-\frac{6}{5} \kappa_S \Big(20 \kappa_{S}^{2} \kappa_{S}^{*,2} +20 \lambda_S |\kappa_S|^2 \lambda_S^* \nonumber \\ 
 &+|\lambda_S|^2 \Big(10 \lambda_S \lambda_S^*  -15 g_{2}^{2}  + 15 \mbox{Tr}\Big({Y_d  Y_{d}^{\dagger}}\Big)  + 15 \mbox{Tr}\Big({Y_u  Y_{u}^{\dagger}}\Big)  + 30 \lambda_T \lambda_T^*  -3 g_{1}^{2}  + 5 \mbox{Tr}\Big({Y_e  Y_{e}^{\dagger}}\Big) \Big)\Big)
\end{align}} 
\subsection{Bilinear Superpotential Parameters}
{\allowdisplaybreaks  \begin{align} 
\beta_{\mu}^{(1)} & =  
2 \mu |\lambda_S|^2  -3 g_{2}^{2} \mu  + 3 \mu \mbox{Tr}\Big({Y_d  Y_{d}^{\dagger}}\Big)  + 3 \mu \mbox{Tr}\Big({Y_u  Y_{u}^{\dagger}}\Big)  + 6 \mu |\lambda_T|^2  -\frac{3}{5} g_{1}^{2} \mu  + \mu \mbox{Tr}\Big({Y_e  Y_{e}^{\dagger}}\Big) \\ 
\beta_{\mu}^{(2)} & =  
+\frac{207}{50} g_{1}^{4} \mu +\frac{9}{5} g_{1}^{2} g_{2}^{2} \mu +\frac{27}{2} g_{2}^{4} \mu +24 g_{2}^{2} \mu |\lambda_T|^2 -4 \lambda_S \mu |\kappa_S|^2 \lambda_S^* -6 \lambda_{S}^{2} \mu \lambda_{S}^{*,2} \nonumber \\ 
 &-30 \lambda_{T}^{2} \mu \lambda_{T}^{*,2} -\frac{2}{5} g_{1}^{2} \mu \mbox{Tr}\Big({Y_d  Y_{d}^{\dagger}}\Big) +16 g_{3}^{2} \mu \mbox{Tr}\Big({Y_d  Y_{d}^{\dagger}}\Big) -9 \mu |\lambda_T|^2 \mbox{Tr}\Big({Y_d  Y_{d}^{\dagger}}\Big) \nonumber \\ 
 &+\frac{6}{5} g_{1}^{2} \mu \mbox{Tr}\Big({Y_e  Y_{e}^{\dagger}}\Big) -3 \mu |\lambda_T|^2 \mbox{Tr}\Big({Y_e  Y_{e}^{\dagger}}\Big) +\frac{4}{5} g_{1}^{2} \mu \mbox{Tr}\Big({Y_u  Y_{u}^{\dagger}}\Big) +16 g_{3}^{2} \mu \mbox{Tr}\Big({Y_u  Y_{u}^{\dagger}}\Big) \nonumber \\ 
 &-9 \mu |\lambda_T|^2 \mbox{Tr}\Big({Y_u  Y_{u}^{\dagger}}\Big) - \mu |\lambda_S|^2 \Big(12 \lambda_T \lambda_T^*  + 3 \mbox{Tr}\Big({Y_d  Y_{d}^{\dagger}}\Big)  + 3 \mbox{Tr}\Big({Y_u  Y_{u}^{\dagger}}\Big)  + \mbox{Tr}\Big({Y_e  Y_{e}^{\dagger}}\Big)\Big)\nonumber \\ 
 &-9 \mu \mbox{Tr}\Big({Y_d  Y_{d}^{\dagger}  Y_d  Y_{d}^{\dagger}}\Big) -6 \mu \mbox{Tr}\Big({Y_d  Y_{u}^{\dagger}  Y_u  Y_{d}^{\dagger}}\Big) -3 \mu \mbox{Tr}\Big({Y_e  Y_{e}^{\dagger}  Y_e  Y_{e}^{\dagger}}\Big) -9 \mu \mbox{Tr}\Big({Y_u  Y_{u}^{\dagger}  Y_u  Y_{u}^{\dagger}}\Big) \\ 
\beta_{M_T}^{(1)} & =  
4 M_T |\lambda_T|^2  -8 g_{2}^{2} M_T \\ 
\beta_{M_T}^{(2)} & =  
\frac{4}{5} M_T \Big(70 g_{2}^{4} -30 \lambda_{T}^{2} \lambda_{T}^{*,2} \nonumber \\ 
 &+|\lambda_T|^2 \Big(-10 \lambda_S \lambda_S^*  -15 \mbox{Tr}\Big({Y_d  Y_{d}^{\dagger}}\Big)  -15 \mbox{Tr}\Big({Y_u  Y_{u}^{\dagger}}\Big)  + 3 g_{1}^{2}  -5 g_{2}^{2}  -5 \mbox{Tr}\Big({Y_e  Y_{e}^{\dagger}}\Big) \Big)\Big)\\ 
\beta_{M_O}^{(1)} & =  
-12 g_{3}^{2} M_O \\ 
\beta_{M_O}^{(2)} & =  
72 g_{3}^{4} M_O \\ 
\beta_{M_S}^{(1)} & =  
4 M_S \Big(|\kappa_S|^2 + |\lambda_S|^2\Big)\\ 
\beta_{M_S}^{(2)} & =  
-\frac{4}{5} M_S \Big(20 \kappa_{S}^{2} \kappa_{S}^{*,2} +20 \lambda_S |\kappa_S|^2 \lambda_S^* \nonumber \\ 
 &+|\lambda_S|^2 \Big(10 \lambda_S \lambda_S^*  -15 g_{2}^{2}  + 15 \mbox{Tr}\Big({Y_d  Y_{d}^{\dagger}}\Big)  + 15 \mbox{Tr}\Big({Y_u  Y_{u}^{\dagger}}\Big)  + 30 \lambda_T \lambda_T^*  -3 g_{1}^{2}  + 5 \mbox{Tr}\Big({Y_e  Y_{e}^{\dagger}}\Big) \Big)\Big)
\end{align}} 
\subsection{Linear Superpotential Parameters}
{\allowdisplaybreaks  \begin{align} 
\beta_{L_S}^{(1)} & =  
2 L_S \Big(|\kappa_S|^2 + |\lambda_S|^2\Big)\\ 
\beta_{L_S}^{(2)} & =  
-\frac{2}{5} L_S \Big(20 \kappa_{S}^{2} \kappa_{S}^{*,2} +20 \lambda_S |\kappa_S|^2 \lambda_S^* \nonumber \\ 
 &+|\lambda_S|^2 \Big(10 \lambda_S \lambda_S^*  -15 g_{2}^{2}  + 15 \mbox{Tr}\Big({Y_d  Y_{d}^{\dagger}}\Big)  + 15 \mbox{Tr}\Big({Y_u  Y_{u}^{\dagger}}\Big)  + 30 \lambda_T \lambda_T^*  -3 g_{1}^{2}  + 5 \mbox{Tr}\Big({Y_e  Y_{e}^{\dagger}}\Big) \Big)\Big)
\end{align}} 
\subsection{Trilinear Soft-Breaking Parameters}
{\allowdisplaybreaks
} 
\subsection{Bilinear Soft-Breaking Parameters}
{\allowdisplaybreaks  \begin{align} 
\beta_{B_{\mu}}^{(1)} & =  
+\frac{6}{5} g_{1}^{2} M_1 \mu +6 g_{2}^{2} M_2 \mu +2 \lambda_S B_S \kappa_S^* +4 \mu \lambda_S^* A_S +12 \mu \lambda_T^* A_T \nonumber \\ 
 &+B_{\mu} \Big(-3 g_{2}^{2}  + 3 \mbox{Tr}\Big({Y_d  Y_{d}^{\dagger}}\Big)  + 3 \mbox{Tr}\Big({Y_u  Y_{u}^{\dagger}}\Big)  + 6 |\lambda_S|^2  + 6 |\lambda_T|^2  -\frac{3}{5} g_{1}^{2}  + \mbox{Tr}\Big({Y_e  Y_{e}^{\dagger}}\Big)\Big)+6 \mu \mbox{Tr}\Big({Y_{d}^{\dagger}  A_d}\Big) \nonumber \\ 
 &+2 \mu \mbox{Tr}\Big({Y_{e}^{\dagger}  A_e}\Big) +6 \mu \mbox{Tr}\Big({Y_{u}^{\dagger}  A_u}\Big) \\ 
\beta_{B_{\mu}}^{(2)} & =  
+B_{\mu} \Big(\frac{207}{50} g_{1}^{4} +\frac{9}{5} g_{1}^{2} g_{2}^{2} +\frac{27}{2} g_{2}^{4} -14 \lambda_{S}^{2} \lambda_{S}^{*,2} -30 \lambda_{T}^{2} \lambda_{T}^{*,2} -\frac{2}{5} g_{1}^{2} \mbox{Tr}\Big({Y_d  Y_{d}^{\dagger}}\Big) +16 g_{3}^{2} \mbox{Tr}\Big({Y_d  Y_{d}^{\dagger}}\Big) \nonumber \\ 
 &+\frac{6}{5} g_{1}^{2} \mbox{Tr}\Big({Y_e  Y_{e}^{\dagger}}\Big) \nonumber \\ 
 &+\frac{1}{5} |\lambda_S|^2 \Big(180 g_{2}^{2}  -180 \lambda_T \lambda_T^*  -20 \kappa_S \kappa_S^*  -25 \mbox{Tr}\Big({Y_e  Y_{e}^{\dagger}}\Big)  + 36 g_{1}^{2}  -75 \mbox{Tr}\Big({Y_d  Y_{d}^{\dagger}}\Big)  -75 \mbox{Tr}\Big({Y_u  Y_{u}^{\dagger}}\Big) \Big)\nonumber \\ 
 &+3 |\lambda_T|^2 \Big(-3 \mbox{Tr}\Big({Y_d  Y_{d}^{\dagger}}\Big)  -3 \mbox{Tr}\Big({Y_u  Y_{u}^{\dagger}}\Big)  + 8 g_{2}^{2}  - \mbox{Tr}\Big({Y_e  Y_{e}^{\dagger}}\Big) \Big)+\frac{4}{5} g_{1}^{2} \mbox{Tr}\Big({Y_u  Y_{u}^{\dagger}}\Big) \nonumber \\ 
 &+16 g_{3}^{2} \mbox{Tr}\Big({Y_u  Y_{u}^{\dagger}}\Big) -9 \mbox{Tr}\Big({Y_d  Y_{d}^{\dagger}  Y_d  Y_{d}^{\dagger}}\Big) -6 \mbox{Tr}\Big({Y_d  Y_{u}^{\dagger}  Y_u  Y_{d}^{\dagger}}\Big) -3 \mbox{Tr}\Big({Y_e  Y_{e}^{\dagger}  Y_e  Y_{e}^{\dagger}}\Big) -9 \mbox{Tr}\Big({Y_u  Y_{u}^{\dagger}  Y_u  Y_{u}^{\dagger}}\Big) \Big)\nonumber \\ 
 &-\frac{2}{25} \Big(207 g_{1}^{4} M_1 \mu +45 g_{1}^{2} g_{2}^{2} M_1 \mu +45 g_{1}^{2} g_{2}^{2} M_2 \mu +675 g_{2}^{4} M_2 \mu +600 g_{2}^{2} M_2 \mu |\lambda_T|^2 \nonumber \\ 
 &+100 \lambda_S \Big(|\kappa_S|^2 + |\lambda_S|^2\Big)B_S \kappa_S^* +100 M_S \lambda_S \kappa_{S}^{*,2} A_\kappa +400 \lambda_S \mu \lambda_{S}^{*,2} A_S -600 g_{2}^{2} \mu \lambda_T^* A_T \nonumber \\ 
 &+1500 \lambda_T \mu \lambda_{T}^{*,2} A_T -10 g_{1}^{2} M_1 \mu \mbox{Tr}\Big({Y_d  Y_{d}^{\dagger}}\Big) +400 g_{3}^{2} M_3 \mu \mbox{Tr}\Big({Y_d  Y_{d}^{\dagger}}\Big) +225 \mu \lambda_T^* A_T \mbox{Tr}\Big({Y_d  Y_{d}^{\dagger}}\Big) \nonumber \\ 
 &+30 g_{1}^{2} M_1 \mu \mbox{Tr}\Big({Y_e  Y_{e}^{\dagger}}\Big) +75 \mu \lambda_T^* A_T \mbox{Tr}\Big({Y_e  Y_{e}^{\dagger}}\Big) +20 g_{1}^{2} M_1 \mu \mbox{Tr}\Big({Y_u  Y_{u}^{\dagger}}\Big) \nonumber \\ 
 &+400 g_{3}^{2} M_3 \mu \mbox{Tr}\Big({Y_u  Y_{u}^{\dagger}}\Big) +225 \mu \lambda_T^* A_T \mbox{Tr}\Big({Y_u  Y_{u}^{\dagger}}\Big) +10 g_{1}^{2} \mu \mbox{Tr}\Big({Y_{d}^{\dagger}  A_d}\Big) -400 g_{3}^{2} \mu \mbox{Tr}\Big({Y_{d}^{\dagger}  A_d}\Big) \nonumber \\ 
 &+225 \mu |\lambda_T|^2 \mbox{Tr}\Big({Y_{d}^{\dagger}  A_d}\Big) -30 g_{1}^{2} \mu \mbox{Tr}\Big({Y_{e}^{\dagger}  A_e}\Big) +75 \mu |\lambda_T|^2 \mbox{Tr}\Big({Y_{e}^{\dagger}  A_e}\Big) -20 g_{1}^{2} \mu \mbox{Tr}\Big({Y_{u}^{\dagger}  A_u}\Big) \nonumber \\ 
 &-400 g_{3}^{2} \mu \mbox{Tr}\Big({Y_{u}^{\dagger}  A_u}\Big) +225 \mu |\lambda_T|^2 \mbox{Tr}\Big({Y_{u}^{\dagger}  A_u}\Big) \nonumber \\ 
 &+5 \lambda_S^* \Big(20 \kappa_S^* \Big(\Big(\kappa_S \mu  + M_S \lambda_S \Big)A_S  + \lambda_S \mu A_\kappa \Big)\nonumber \\ 
 &+\mu \Big(60 \lambda_T^* \Big(2 \lambda_S A_T  + \lambda_T A_S \Big)+5 A_S \Big(3 \mbox{Tr}\Big({Y_d  Y_{d}^{\dagger}}\Big)  + 3 \mbox{Tr}\Big({Y_u  Y_{u}^{\dagger}}\Big)  + \mbox{Tr}\Big({Y_e  Y_{e}^{\dagger}}\Big)\Big)\nonumber \\ 
 &+3 \lambda_S \Big(15 \mbox{Tr}\Big({Y_{d}^{\dagger}  A_d}\Big)  + 15 \mbox{Tr}\Big({Y_{u}^{\dagger}  A_u}\Big)  + 30 g_{2}^{2} M_2  + 5 \mbox{Tr}\Big({Y_{e}^{\dagger}  A_e}\Big)  + 6 g_{1}^{2} M_1 \Big)\Big)\Big)\nonumber \\ 
 &+450 \mu \mbox{Tr}\Big({Y_d  Y_{d}^{\dagger}  A_d  Y_{d}^{\dagger}}\Big) +150 \mu \mbox{Tr}\Big({Y_d  Y_{u}^{\dagger}  A_u  Y_{d}^{\dagger}}\Big) +150 \mu \mbox{Tr}\Big({Y_e  Y_{e}^{\dagger}  A_e  Y_{e}^{\dagger}}\Big) \nonumber \\ 
 &+150 \mu \mbox{Tr}\Big({Y_u  Y_{d}^{\dagger}  A_d  Y_{u}^{\dagger}}\Big) +450 \mu \mbox{Tr}\Big({Y_u  Y_{u}^{\dagger}  A_u  Y_{u}^{\dagger}}\Big) \Big)\\ 
\beta_{B_T}^{(1)} & =  
4 \Big(-2 g_{2}^{2} B_T  + 2 M_T \lambda_T^* A_T  + 4 g_{2}^{2} M_2 M_T  + |\lambda_T|^2 B_T \Big)\\ 
\beta_{B_T}^{(2)} & =  
-\frac{4}{5} \Big(B_T \Big(30 \lambda_{T}^{2} \lambda_{T}^{*,2}  -70 g_{2}^{4}  \nn\\
&+ |\lambda_T|^2 \Big(10 \lambda_S \lambda_S^*  + 15 \mbox{Tr}\Big({Y_d  Y_{d}^{\dagger}}\Big)  + 15 \mbox{Tr}\Big({Y_u  Y_{u}^{\dagger}}\Big)  -3 g_{1}^{2}  + 5 g_{2}^{2}  + 5 \mbox{Tr}\Big({Y_e  Y_{e}^{\dagger}}\Big) \Big)\Big)\nonumber \\ 
 &+2 M_T \Big(140 g_{2}^{4} M_2 +60 \lambda_T \lambda_{T}^{*,2} A_T \nonumber \\ 
 &+\lambda_T^* \Big(10 \lambda_S^* \Big(\lambda_S A_T  + \lambda_T A_S \Big)+A_T \Big(15 \mbox{Tr}\Big({Y_d  Y_{d}^{\dagger}}\Big)  + 15 \mbox{Tr}\Big({Y_u  Y_{u}^{\dagger}}\Big)  -3 g_{1}^{2}  + 5 g_{2}^{2}  + 5 \mbox{Tr}\Big({Y_e  Y_{e}^{\dagger}}\Big) \Big)\nonumber \\ 
 &+\lambda_T \Big(15 \mbox{Tr}\Big({Y_{d}^{\dagger}  A_d}\Big)  + 15 \mbox{Tr}\Big({Y_{u}^{\dagger}  A_u}\Big)  + 3 g_{1}^{2} M_1  -5 g_{2}^{2} M_2  + 5 \mbox{Tr}\Big({Y_{e}^{\dagger}  A_e}\Big) \Big)\Big)\Big)\Big)\\ 
\beta_{B_O}^{(1)} & =  
12 g_{3}^{2} \Big(2 M_3 M_O  - B_O \Big)\\ 
\beta_{B_O}^{(2)} & =  
72 g_{3}^{4} \Big(-4 M_3 M_O  + B_O\Big)\\ 
\beta_{B_S}^{(1)} & =  
4 \Big(2 |\kappa_S|^2  + |\lambda_S|^2\Big)B_S  + 8 \Big(\kappa_S B_{\mu} \lambda_S^*  + M_S \kappa_S^* A_\kappa  + M_S \lambda_S^* A_S \Big)\\ 
\beta_{B_S}^{(2)} & =  
-\frac{4}{5} \Big(B_S \Big(40 \kappa_{S}^{2} \kappa_{S}^{*,2} +40 \lambda_S |\kappa_S|^2 \lambda_S^* \nonumber \\ 
 &+|\lambda_S|^2 \Big(10 \lambda_S \lambda_S^*  -15 g_{2}^{2}  + 15 \mbox{Tr}\Big({Y_d  Y_{d}^{\dagger}}\Big)  + 15 \mbox{Tr}\Big({Y_u  Y_{u}^{\dagger}}\Big)  + 30 \lambda_T \lambda_T^*  -3 g_{1}^{2}  + 5 \mbox{Tr}\Big({Y_e  Y_{e}^{\dagger}}\Big) \Big)\Big)\nonumber \\ 
 &+2 \Big(50 M_S \kappa_S \kappa_{S}^{*,2} A_\kappa +10 \lambda_{S}^{*,2} \Big(\Big(2 M_S \lambda_S  + \kappa_S \mu \Big)A_S  + \kappa_S \lambda_S B_{\mu} \Big)\nonumber \\ 
 &+\lambda_S^* \Big(3 g_{1}^{2} M_1 M_S \lambda_S +15 g_{2}^{2} M_2 M_S \lambda_S +9 g_{1}^{2} M_1 \kappa_S \mu +45 g_{2}^{2} M_2 \kappa_S \mu -3 g_{1}^{2} M_S A_S -15 g_{2}^{2} M_S A_S \nonumber \\ 
 &+30 M_S |\lambda_T|^2 A_S +10 M_S \kappa_S^* \Big(2 \lambda_S A_\kappa  + 3 \kappa_S A_S \Big)+30 M_S \lambda_S \lambda_T^* A_T +30 \kappa_S \mu \lambda_T^* A_T \nn\\
&+15 M_S A_S \mbox{Tr}\Big({Y_d  Y_{d}^{\dagger}}\Big) \nonumber \\ 
 &+5 M_S A_S \mbox{Tr}\Big({Y_e  Y_{e}^{\dagger}}\Big) +15 M_S A_S \mbox{Tr}\Big({Y_u  Y_{u}^{\dagger}}\Big) \nonumber \\ 
 &+\kappa_S B_{\mu} \Big(15 \mbox{Tr}\Big({Y_d  Y_{d}^{\dagger}}\Big)  + 15 \mbox{Tr}\Big({Y_u  Y_{u}^{\dagger}}\Big)  + 30 |\lambda_T|^2  -45 g_{2}^{2}  + 5 \mbox{Tr}\Big({Y_e  Y_{e}^{\dagger}}\Big)  -9 g_{1}^{2} \Big)+15 M_S \lambda_S \mbox{Tr}\Big({Y_{d}^{\dagger}  A_d}\Big) \nonumber \\ 
 &+15 \kappa_S \mu \mbox{Tr}\Big({Y_{d}^{\dagger}  A_d}\Big) +5 M_S \lambda_S \mbox{Tr}\Big({Y_{e}^{\dagger}  A_e}\Big) +5 \kappa_S \mu \mbox{Tr}\Big({Y_{e}^{\dagger}  A_e}\Big) +15 M_S \lambda_S \mbox{Tr}\Big({Y_{u}^{\dagger}  A_u}\Big) \nonumber \\ 
 &+15 \kappa_S \mu \mbox{Tr}\Big({Y_{u}^{\dagger}  A_u}\Big) \Big)\Big)\Big)
\end{align}} 
\subsection{Linear Soft-Breaking Parameters}
{\allowdisplaybreaks  \begin{align} 
(4\pi)^2 X_S^{(1)} & =  
2 \Big(2 m_S^2 \kappa_S M_S^* +M_S B_S \kappa_S^* +2 M_S B_{\mu} \lambda_S^* +2 m_{H_d}^2 \lambda_S \mu^* +2 m_{H_u}^2 \lambda_S \mu^* +|\kappa_S|^2 t_S +|\lambda_S|^2 t_S \nn\\&+2 L_S \kappa_S^* A_\kappa +B_S^* A_\kappa +2 L_S \lambda_S^* A_S +2 B_{\mu}^* A_S \Big)\\ 
(4\pi)^4 X_S^{(2)} & =  
-\frac{2}{5} \Big(6 g_{1}^{2} L_S M_1 |\lambda_S|^2 +30 g_{2}^{2} L_S M_2 |\lambda_S|^2 +20 M_S \Big(|\kappa_S|^2 + |\lambda_S|^2\Big)B_S \kappa_S^* +6 g_{1}^{2} M_1 M_S \mu \lambda_S^* \nonumber \\ 
 &+30 g_{2}^{2} M_2 M_S \mu \lambda_S^* -6 g_{1}^{2} M_S B_{\mu} \lambda_S^* -30 g_{2}^{2} M_S B_{\mu} \lambda_S^* +60 M_S |\lambda_T|^2 B_{\mu} \lambda_S^* +20 M_S \lambda_S B_{\mu} \lambda_{S}^{*,2} \nonumber \\ 
 &-6 g_{1}^{2} m_{H_d}^2 \lambda_S \mu^* -30 g_{2}^{2} m_{H_d}^2 \lambda_S \mu^* -6 g_{1}^{2} m_{H_u}^2 \lambda_S \mu^* -30 g_{2}^{2} m_{H_u}^2 \lambda_S \mu^* \nonumber \\ 
 &-12 g_{1}^{2} \lambda_S |M_1|^2 \mu^* -60 g_{2}^{2} \lambda_S |M_2|^2 \mu^* +120 m_{H_d}^2 \lambda_S |\lambda_T|^2 \mu^* +120 m_{H_u}^2 \lambda_S |\lambda_T|^2 \mu^* \nonumber \\ 
 &+60 m_T^2 \lambda_S |\lambda_T|^2 \mu^* +40 \lambda_S |A_S|^2 \mu^* +60 \lambda_S |A_T|^2 \mu^* +40 m_{H_d}^2 \lambda_{S}^{2} \lambda_S^* \mu^* +40 m_{H_u}^2 \lambda_{S}^{2} \lambda_S^* \mu^* \nonumber \\ 
 &+20 m_S^2 \lambda_{S}^{2} \lambda_S^* \mu^* +6 g_{1}^{2} M_1 \lambda_S B_{\mu}^* +30 g_{2}^{2} M_2 \lambda_S B_{\mu}^* -3 g_{1}^{2} |\lambda_S|^2 t_S -15 g_{2}^{2} |\lambda_S|^2 t_S \nonumber \\ 
 &+20 \kappa_{S}^{2} \kappa_{S}^{*,2} t_S +20 \lambda_S |\kappa_S|^2 \lambda_S^* t_S +10 \lambda_{S}^{2} \lambda_{S}^{*,2} t_S +30 \lambda_T |\lambda_S|^2 \lambda_T^* t_S +40 L_S |\lambda_S|^2 \kappa_S^* A_\kappa \nonumber \\ 
 &+20 M_{S}^{2} \kappa_{S}^{*,2} A_\kappa +80 L_S \kappa_S \kappa_{S}^{*,2} A_\kappa +40 |\kappa_S|^2 B_S^* A_\kappa +20 |\lambda_S|^2 B_S^* A_\kappa \nonumber \\ 
 &+20 M_S^* \Big(2 \kappa_S |A_\kappa|^2  + \Big(3 m_S^2  + m_{H_d}^2 + m_{H_u}^2\Big)\kappa_S |\lambda_S|^2  + 5 m_S^2 \kappa_{S}^{2} \kappa_S^*  + \kappa_S |A_S|^2  + \lambda_S A_S^* A_\kappa \Big)\nonumber \\ 
 &-6 g_{1}^{2} L_S \lambda_S^* A_S -30 g_{2}^{2} L_S \lambda_S^* A_S +40 L_S |\kappa_S|^2 \lambda_S^* A_S +60 L_S |\lambda_T|^2 \lambda_S^* A_S +20 M_{S}^{2} \kappa_S^* \lambda_S^* A_S \nonumber \\ 
 &+40 L_S \lambda_S \lambda_{S}^{*,2} A_S +20 M_S \mu \lambda_{S}^{*,2} A_S +6 g_{1}^{2} M_1 \mu^* A_S +30 g_{2}^{2} M_2 \mu^* A_S +20 \kappa_S \lambda_S^* B_S^* A_S \nonumber \\ 
 &-6 g_{1}^{2} B_{\mu}^* A_S -30 g_{2}^{2} B_{\mu}^* A_S +40 |\lambda_S|^2 B_{\mu}^* A_S +60 |\lambda_T|^2 B_{\mu}^* A_S +60 \lambda_T \mu^* A_T^* A_S \nonumber \\ 
 &+60 L_S |\lambda_S|^2 \lambda_T^* A_T +60 M_S \mu \lambda_S^* \lambda_T^* A_T +60 \lambda_S \lambda_T^* B_{\mu}^* A_T +30 M_S B_{\mu} \lambda_S^* \mbox{Tr}\Big({Y_d  Y_{d}^{\dagger}}\Big) \nonumber \\ 
 &+60 m_{H_d}^2 \lambda_S \mu^* \mbox{Tr}\Big({Y_d  Y_{d}^{\dagger}}\Big) +30 m_{H_u}^2 \lambda_S \mu^* \mbox{Tr}\Big({Y_d  Y_{d}^{\dagger}}\Big) +15 |\lambda_S|^2 t_S \mbox{Tr}\Big({Y_d  Y_{d}^{\dagger}}\Big) \nonumber \\ 
 &+30 L_S \lambda_S^* A_S \mbox{Tr}\Big({Y_d  Y_{d}^{\dagger}}\Big) +30 B_{\mu}^* A_S \mbox{Tr}\Big({Y_d  Y_{d}^{\dagger}}\Big) +10 M_S B_{\mu} \lambda_S^* \mbox{Tr}\Big({Y_e  Y_{e}^{\dagger}}\Big) \nonumber \\ 
 &+20 m_{H_d}^2 \lambda_S \mu^* \mbox{Tr}\Big({Y_e  Y_{e}^{\dagger}}\Big) +10 m_{H_u}^2 \lambda_S \mu^* \mbox{Tr}\Big({Y_e  Y_{e}^{\dagger}}\Big) +5 |\lambda_S|^2 t_S \mbox{Tr}\Big({Y_e  Y_{e}^{\dagger}}\Big) \nonumber \\ 
 &+10 L_S \lambda_S^* A_S \mbox{Tr}\Big({Y_e  Y_{e}^{\dagger}}\Big) +10 B_{\mu}^* A_S \mbox{Tr}\Big({Y_e  Y_{e}^{\dagger}}\Big) +30 M_S B_{\mu} \lambda_S^* \mbox{Tr}\Big({Y_u  Y_{u}^{\dagger}}\Big) \nonumber \\ 
 &+30 m_{H_d}^2 \lambda_S \mu^* \mbox{Tr}\Big({Y_u  Y_{u}^{\dagger}}\Big) +60 m_{H_u}^2 \lambda_S \mu^* \mbox{Tr}\Big({Y_u  Y_{u}^{\dagger}}\Big) +15 |\lambda_S|^2 t_S \mbox{Tr}\Big({Y_u  Y_{u}^{\dagger}}\Big) \nonumber \\ 
 &+30 L_S \lambda_S^* A_S \mbox{Tr}\Big({Y_u  Y_{u}^{\dagger}}\Big) +30 B_{\mu}^* A_S \mbox{Tr}\Big({Y_u  Y_{u}^{\dagger}}\Big) +30 L_S |\lambda_S|^2 \mbox{Tr}\Big({Y_{d}^{\dagger}  A_d}\Big) +30 M_S \mu \lambda_S^* \mbox{Tr}\Big({Y_{d}^{\dagger}  A_d}\Big) \nonumber \\ 
 &+30 \lambda_S B_{\mu}^* \mbox{Tr}\Big({Y_{d}^{\dagger}  A_d}\Big) +10 L_S |\lambda_S|^2 \mbox{Tr}\Big({Y_{e}^{\dagger}  A_e}\Big) +10 M_S \mu \lambda_S^* \mbox{Tr}\Big({Y_{e}^{\dagger}  A_e}\Big) +10 \lambda_S B_{\mu}^* \mbox{Tr}\Big({Y_{e}^{\dagger}  A_e}\Big) \nonumber \\ 
 &+30 L_S |\lambda_S|^2 \mbox{Tr}\Big({Y_{u}^{\dagger}  A_u}\Big) +30 M_S \mu \lambda_S^* \mbox{Tr}\Big({Y_{u}^{\dagger}  A_u}\Big) +30 \lambda_S B_{\mu}^* \mbox{Tr}\Big({Y_{u}^{\dagger}  A_u}\Big) +30 \mu^* A_S \mbox{Tr}\Big({A_d^*  Y_{d}^{T}}\Big) \nonumber \\ 
 &+30 \lambda_S \mu^* \mbox{Tr}\Big({A_d^*  A_{d}^{T}}\Big) +10 \mu^* A_S \mbox{Tr}\Big({A_e^*  Y_{e}^{T}}\Big) +10 \lambda_S \mu^* \mbox{Tr}\Big({A_e^*  A_{e}^{T}}\Big) +30 \mu^* A_S \mbox{Tr}\Big({A_u^*  Y_{u}^{T}}\Big) \nonumber \\ 
 &+30 \lambda_S \mu^* \mbox{Tr}\Big({A_u^*  A_{u}^{T}}\Big) +30 \lambda_S \mu^* \mbox{Tr}\Big({Y_d  Y_{d}^{\dagger}  m_d^{2 *}}\Big) +30 \lambda_S \mu^* \mbox{Tr}\Big({Y_d  m_q^{2 *}  Y_{d}^{\dagger}}\Big) \nonumber \\ 
 &+10 \lambda_S \mu^* \mbox{Tr}\Big({Y_e  Y_{e}^{\dagger}  m_e^{2 *}}\Big) +10 \lambda_S \mu^* \mbox{Tr}\Big({Y_e  m_l^{2 *}  Y_{e}^{\dagger}}\Big) +30 \lambda_S \mu^* \mbox{Tr}\Big({Y_u  Y_{u}^{\dagger}  m_u^{2 *}}\Big) \nonumber \\ 
 &+30 \lambda_S \mu^* \mbox{Tr}\Big({Y_u  m_q^{2 *}  Y_{u}^{\dagger}}\Big) \Big)
\label{EQ:STANDARDTADPOLES}\end{align}} 
\subsection{Soft-Breaking Scalar Masses}
\begin{align} 
\sigma_{1,1} & = \sqrt{\frac{3}{5}} g_1 \Big(-2 \mbox{Tr}\Big({m_u^2}\Big)  - m_{H_d}^2  - \tr\Big(m_l^2\Big)\  + m_{H_u}^2 + \tr\Big(m_q^2\Big) + \mbox{Tr}\Big({m_d^2}\Big) + \mbox{Tr}\Big({m_e^2}\Big)\Big)\nn\\ 
\text{Tr2U1}\Big(1,1\Big) & = \frac{1}{10} g_{1}^{2} \Big(2 \mbox{Tr}\Big({m_d^2}\Big)  + 3 m_{H_d}^2  + 3 m_{H_u}^2  + 3 \tr\Big(m_l^2\Big)  + 6 \mbox{Tr}\Big({m_e^2}\Big)  + 8 \mbox{Tr}\Big({m_u^2}\Big)  + \tr\Big(m_q^2\Big)\Big)\nn\\ 
\sigma_{3,1} & = \frac{1}{20} \frac{1}{\sqrt{15}} g_1 \Big(-9 g_{1}^{2} m_{H_d}^2 -45 g_{2}^{2} m_{H_d}^2 +9 g_{1}^{2} m_{H_u}^2 +45 g_{2}^{2} m_{H_u}^2\nn\\
& +30 \Big(- m_{H_u}^2  + m_{H_d}^2\Big)|\lambda_S|^2 +90 \Big(- m_{H_u}^2  + m_{H_d}^2\Big)|\lambda_T|^2 \nonumber \\ 
 &-9 g_{1}^{2} \tr\Big(m_l^2\Big) -45 g_{2}^{2} \tr\Big(m_l^2\Big) +g_{1}^{2} \tr\Big(m_q^2\Big) +45 g_{2}^{2} \tr\Big(m_q^2\Big) +80 g_{3}^{2}\tr\Big(m_q^2\Big) +4 g_{1}^{2} \mbox{Tr}\Big({m_d^2}\Big) \nonumber \\ 
 &+80 g_{3}^{2} \mbox{Tr}\Big({m_d^2}\Big) +36 g_{1}^{2} \mbox{Tr}\Big({m_e^2}\Big) -32 g_{1}^{2} \mbox{Tr}\Big({m_u^2}\Big) -160 g_{3}^{2} \mbox{Tr}\Big({m_u^2}\Big) +90 m_{H_d}^2 \mbox{Tr}\Big({Y_d  Y_{d}^{\dagger}}\Big) \nonumber \\ 
 &+30 m_{H_d}^2 \mbox{Tr}\Big({Y_e  Y_{e}^{\dagger}}\Big) -90 m_{H_u}^2 \mbox{Tr}\Big({Y_u  Y_{u}^{\dagger}}\Big) -60 \mbox{Tr}\Big({Y_d  Y_{d}^{\dagger}  m_d^{2 *}}\Big) -30 \mbox{Tr}\Big({Y_d  m_q^{2 *}  Y_{d}^{\dagger}}\Big) \nonumber \\ 
 &-60 \mbox{Tr}\Big({Y_e  Y_{e}^{\dagger}  m_e^{2 *}}\Big) +30 \mbox{Tr}\Big({Y_e  m_l^{2 *}  Y_{e}^{\dagger}}\Big) +120 \mbox{Tr}\Big({Y_u  Y_{u}^{\dagger}  m_u^{2 *}}\Big) -30 \mbox{Tr}\Big({Y_u  m_q^{2 *}  Y_{u}^{\dagger}}\Big) \Big)\nn\\ 
\sigma_{2,2} & = \frac{1}{2} \Big(3 \tr\Big(m_q^2\Big)  + 6 m_T^2  + m_{H_d}^2 + m_{H_u}^2 + \tr\Big(m_l^2\Big)\Big)\nn\\ 
\sigma_{2,3} & = \frac{1}{2} \Big(16 m_O^2  + 2 \tr \Big(m_q^2\Big)  + \mbox{Tr}\Big({m_d^2}\Big) + \mbox{Tr}\Big({m_u^2}\Big)\Big)
\label{EQ:SIGMAS}\end{align} 
{\allowdisplaybreaks
}

\end{document}